\documentclass[iop,usenatbib]{emulateapj}

\usepackage{epsfig}
\usepackage{subfigure}
\usepackage{multirow}

\def\HI{\hbox{H\hskip1.5pt$\scriptstyle\rm I\ $}}
\def\HIp{\hbox{H\hskip1.5pt$\scriptstyle\rm I$}}

\def\HII{\hbox{H\hskip1.5pt$\scriptstyle\rm II\ $}}
\def\HeI{\hbox{He\hskip1.5pt$\scriptstyle\rm I\ $}}

\def\HeII{\hbox{He\hskip1.5pt$\scriptstyle{\rm II}\ $}}
\def\HeIIp{\hbox{He\hskip1.5pt$\scriptstyle{\rm II}$}}
\def\HeIII{\hbox{He\hskip1.5pt$\scriptstyle\rm III\ $}}
\def\HeIIIp{\hbox{He\hskip1.5pt$\scriptstyle\rm III$}}

\def\CIVp{\hbox{C\hskip1.5pt$\scriptstyle\rm IV$}}
\def\nHI{_{\rm HI}}

\def\nHeIII{_{\rm HeIII}}

\def\nHII{_{\rm HII}}
\def\H{_{\rm H}}
\def\He{_{\rm He}}
\def\HH{H$_2$\ }

\def\HHp{H$_2$}

\def\lya{Ly$\alpha$}
\def\lyc{LyC}

\def\es{}

\def\ini{_{\rm start}}

\def\dis{_{\rm dis}}
\def\esc{_{\rm esc}}
\def\trap{_{\rm trap}}
\def\escG{_{\rm esc,G}}
\def\escAGN{_{\rm esc,AGN}}
\def\rec{_{\rm mix}}
\def\der{{\rm d}}

\def\C{_{\rm C}}
\def\B{_{\rm B}}
\def\D{_{\rm D}}
\def\G{_{\rm G}}

\def\lav{\langle}
\def\rav{\rangle}

\def\T{_{\rm T}}
\def\se{^{\rm e}}

\def\AGN{_{\rm AGN}}
\def\SN{_{\rm SN}}
\def\PISN{_{\rm PISN}}

\def\sec{Section\ }

\def\D{_{\rm D}}
\def\B{_{\rm B}}
\def\ii{_{\rm i}}
\def\Si{_{\rm SII}}
\def\bSi{_{\rm \!\!SII}}
\def\I{_{\rm I}}
\def\II{_{\rm II}}
\def\III{_{\rm III}}

\def\minii{_{\rm min,i}}

\def\last{_{\rm last}}

\def\p{_{\rm p}}
\def\ee{_{\rm e}}

\def\J{_{\rm J}}

\def\hg{_{\rm hg}}

\def\ion{_{\rm ion,H}}
\def\ionHe{_{\rm ion,He}}

\def\rhb{_{\rm rh,B}}
\def\rhc{_{\rm rh,C}}
\def\up{^{\rm up}}
\def\lo{^{\rm lo}}

\def\cgb{_{\rm cg,B}}
\def\cgc{_{\rm cg,C}}

\def\SN{_{\rm SN}}

\def\sfc{_{\rm sf,C}}

\def\dynX{_{\rm dyn,C}}

\def\IGM{_{\rm IGM}}
\def\BH{_{\rm BH}}

\def\crit{_{\rm c}}

\def\acc{_{\rm acc}}

\def\u1{^{(1)}}

\def\modot{M$_\odot$}
\def\modotf{{\rm M}_\odot}

\def\pIaII{Pop~II\ }
\def\pIaIId{Pop~II}
\def\pIII{Pop~III\ }
\def\pIIId{Pop~III}

\def\y3{$p_{\rm III}$}
\def\yb3{$p_{\rm III}$\ }
\def\e3{$\epsilon_{\rm III}$}
\def\eb3{$\epsilon_{\rm III}$\ }
\def\beq{\begin{equation}}
\def\eeq{\end{equation}}
\def\beqa{\begin{eqnarray}}
\def\eeqa{\end{eqnarray}}

\begin{document}

\shorttitle{Constraining the Epoch of Reionization}
\shortauthors{Salvador-Sol\'e et al.}

\title{Constraining the epoch of reionization from the
  observed properties\\ of the high-z Universe}

\author{Eduard Salvador-Sol\'e$^1$, Alberto Manrique$^1$, Rafael
  Guzman$^{1,2}$,\\Jos\'e Miguel Rodr\'\i guez Espinosa$^3$, Jes\'us
  Gallego$^4$, Artemio Herrero$^3$,\\J. Miguel Mas-Hesse$^5$, and Antonio
  Mar\'in Franch$^6$}

\affil{$^1$Institut de Ci\`encies del Cosmos. Universitat de Barcelona
  (UB-IEEC), E-08028 Barcelona, Spain\\ $^2$Dept. of
  Astronomy. University of Florida, USA\\ $^3$Instituto de Astrof\'\i sica
  de Canarias y Depto. de Astrof\'\i sica, Univ. de La Laguna,
  E-38205 La Laguna, Spain\\ $^4$Depto. de Astrof\'\i sica, Facultad
  de Físicas. Universidad Complutense de Madrid, E-28040 Madrid,
  Spain\\ $^5$Centro de Astrobiolog\'\i a - Depto. de Astof\'\i sica
  (CSIC-INTA), E-28850 Madrid, Spain\\$^6$Centro de Estudios de F\'\i
  sica del Cosmos de Arag\'on, E-44001 Teruel, Spain}

\email{e.salvador@ub.edu}

\received{2016 March 23}
\revised{2016 October 14}
\accepted{2016 November 5}


\begin{abstract}
\noindent 
We combine observational data on a dozen independent cosmic properties
at high-$z$ with the information on reionization drawn from the
spectra of distant luminous sources and the cosmic microwave
background (CMB) to constrain the interconnected evolution
of galaxies and the intergalactic medium since the dark ages. The
  only acceptable solutions are concentrated in two narrow sets. In
  one of them reionization proceeds in two phases: a first one driven
  by Population III stars, completed at $z\sim 10$, and after a short
  recombination period a second one driven by normal galaxies,
  completed at $z\sim 6$. In the other set both kinds of sources work
  in parallel until full reionization at $z\sim 6$. The best solution
  with double reionization gives excellent fits to all the observed
  cosmic histories, but the CMB optical depth is 3-$\sigma$ larger
  than the recent estimate from the Planck data. Alternatively, the
  best solution with single reionization gives less good fits to the
  observed star formation rate density and cold gas mass density
  histories, but the CMB optical depth is consistent with that
  estimate. We make several predictions, testable with future
  observations, that should discriminate between the two reionization
  scenarios. As a byproduct our models provide a natural explanation
  to some characteristic features of the cosmic properties at
  high-$z$, as well as to the origin of globular clusters.
\end{abstract}


\section{INTRODUCTION}
\label{intro}

Reionization of the cosmic gas after it became essentially neutral at
a redshift $z\sim 1100$ plays a crucial role in the galaxy
formation process. Unfortunately, the details of the epoch of
reionization (EoR) are to a large extent unknown.

The absence of global absorption shortwards of the rest-frame
Lyman-$\alpha$ (\lya) line in the spectra of quasars at $z< 3$ made
\citet{GP65} realize that the hydrogen present in the nearby
intergalactic medium (IGM) was ionized ($\bar x\nHI\la 10^{-4}$)
except for small intervening systems yielding discrete absorption
lines, the so-called \lya\ forest. The Gunn-Peterson trough caused by
neutral intergalactic hydrogen was finally found by \citet{Becker01}
and \citet{DCSM01} in the spectra of quasars at $z\sim 6$, suggesting
that value for the redshift $z\ion$ of full hydrogen ionization.

Several subsequent studies using: (a) the mean opacity of the
\lya\ forest \citep{FEtal06}; (b) the size of the proximity zone
around quasars \citep{WytEtal05,FEtal06,BH07a,LiEtal07,MaEtal07}; (c)
the detection of damping wing absorption by neutral IGM in quasar
spectra \citep{MH04,MH07,MortEtal11b}; (d) its non-detection in the
spectra of gamma ray bursts \citep{ToEtal06,McQEtal07}; (e) the
abundance of \lya\ emitters (LAEs)
\citep{Mal04,HC05,FETal06,McQEtal07,MF08,OEta1O,Jen13a}; and (f) the
covering fraction of dark pixels in the \lya\ and Ly$\beta$ forests
\citep{MMF11} confirmed a value of $z\ion$ of $6.0^{+0.3}_{-0.5}$.

Such a value is not inconsistent with the detection of LAEs beyond
$z=6$. Not only is the frequency of \lya\ photons shifted in origin
due to outflow velocities of the emitting gas clouds, but they are
also sufficiently redshifted when leaving the ionized bubbles around
galaxies for not being absorbed by the neutral hydrogen present in the
IGM. In any event, the LAE abundance rapidly drops beyond $z=6$
\citep{H11}, the most distant spectroscopically confirmed LAE lying at
$z=6.96$ (\citealt{Oea08}; but see \citealt{Zi15}). This would imply
that the volume filling factor of ionized regions, $Q\nHII(z)$, is
decreasing at $z=7$.

An alternative way to estimate $z\ion$ is from the analysis of the
large-scale cosmic microwave background (CMB) anisotropies. Assuming
instantaneous reionization, one can determine $z\ion$ and the optical
depth to Thomson scattering of free electrons by CMB photons,
$\tau\es$ (hereafter simply the CMB optical depth). The 9-year data
gathered from the {\it Wilkinson Microwave Anisotropy Probe} (WMAP9,
\citealt{HinEtal13}) led to $z\ion=10.6 \pm 1.2$ and $\tau\es=0.089\pm
0.014$. The estimates drawn from the temperature three-years {\it
    Planck} data combined with the polarization WMAP data were
  $z\ion=9.9^{+1.8}_{-1.6}$ and $\tau\es=0.078\pm 0.014$ \citep{P15},
  and the most recent ones inferred taking into account the own {\it
    Planck} data on the large-scale polarization anisotropies have
  decreased to $z\ion=8.8^{+0.9}_{-0.8}$ and $\tau\es=0.058\pm 0.012$
  \citep{P16}.

The marked difference between the latter $z\ion$ estimates and the
previous ones drawn from the spectra of distant luminous sources seems
to indicate that reionization, far from being instantaneous, was
extended, and possibly non-monotonous
\citep{WL03,Cen03,CFO03,HH03,HHo03,SYAHS04,NC04,Gnedin04,FL05,WC07,BH07b}. In
this sense, the larger values of $z\ion$ inferred assuming
instantaneous reionization would just indicate that the ionized
fraction was quite low by $z\sim 10$. Of course, the uncertainty in
the associated $\tau\es$ value is also large. However, provided the
ionized fraction were small enough at $z\ga 15$, the value of
$\tau\es$ should not be too biased \citep{Ho03}.

The small-scale CMB anisotropies also inform on the duration $\Delta
z$ of reionization \citep{ZaEtal12,P16}. However, its derivation depends
on the uncertain correction for dust emission, the dominant signal at
the relevant spatial frequencies, and is very sensitive to the assumed
morphological model of ionized bubbles.

Another integral quantity that constrains the EoR, similar to
$\tau\es$ but more sensitive to low redshifts, is the weak
comptonization distortion of the CMB radiation spectrum. Although its
measured upper limit of $y< 1.5\times 10^{-5}$ \citep{Mea90} is a
rather loose constraint, it has the advantage with respect to
$\tau\es$ of being inferred with no modeling.

But the only way to probe the EoR at every $z$ is through 21 cm
hyperfine line observations (see \citealt{MW10} for a comprehensive
review). Unfortunately, the cosmological signal is 5 orders of
magnitude smaller than that produced in galaxies, which greatly
complicates the technique, and makes it necessary to have a previous
good knowledge of the galaxy formation process
\citep{Iea10,Iea08}. Presently, this approach has only allowed one to
put a lower limit to $\Delta z$ of 0.06 \citep{BR10}.

Among the various mechanisms that might contribute to reionization
(see e.g. \citealt{Che07}), the most simple one is photo-ionization by
luminous sources, namely active galactic nuclei (AGN), normal galaxies
with ordinary Population II (\pIaIId) stars, and first generation
metal-free Population III (\pIIId) stars. The X-ray background
demonstrates that AGN emit one order of magnitude less UV photons than
needed at $z\sim 6$ \citep{Worsea05,CEtal09,McQ12}. This implies that
either massive black holes (MBHs) are too scarce \citep{WillEtal10a},
or their associated AGN are too obscured by dust \citep{Trea11}, while
more abundant mini-quasars associated with stellar black holes would
have too short duty cycles \citep{AWA09,MBCO09}. Likewise, the star
formation rate densities derived from the rest-frame UV luminosity
functions (LFs) of normal bright galaxies
\citep{McLEtal10,LorEtal11,BouwEtal11,BouwEtal12a} show that these
objects are insufficient to ionize the IGM by $z \sim 6$. Therefore,
ionization would be achieved either as a result of a substantial
abundance of faint star-forming galaxies \citep{Rea15,I15,BouwEtal15}
or of the still undetected first generation \pIII stars
\citep{SYAHS04,STC08}.

The situation is even more uncertain regarding \HeII reionization. The
\HeII mean opacity inferred from the \lya\ forest suggests that the
redshift of complete \HeII reionization, $z\ionHe$, should be less
than or approximately equal to 3 \citep{RGS00,S00,DF09,McQ09,BBHS11},
and probably no smaller than 2 as the AGN emission begins to decline
at that redshift \citep{CEtal09}. AGN are indeed the most plausible
ionizing sources responsible of \HeII reionization as they emit more
energetic photons than normal galaxies, and \pIII stars, the other
sources of energetic photons, do not form at $z< 6$.

It is thus clear that to fully determine the EoR we must monitor the
formation and evolution of the three kinds of luminous sources and
their feedback since the dark ages. The problem with this approach is
that some aspects of the physics involved are poorly known and must be
treated through numerous parameters. This affects not only
semi-analytic models (SAMs) but also cosmological hydrodynamic
simulations. Indeed, simulations accurately deal with dark matter
(DM), but most of the baryon physics is at sub-resolution
scale. Consequently, they use essentially the same recipes and
parameters as SAMs in this regard. This problem is particularly hard
for simulations because the huge CPU time they involve severely limits
the coverage of the whole parameter space. Thus SAMs are the best
option.
 
The observational data on the Universe at high-$z$ are rapidly
growing, and it might nowadays be possible to address successfully
this problem.  That is, it might be possible to adjust all the free
parameters of a SAM-like model such as the {\it Analytic Model of
  Intergalactic-medium and GAlaxies} (AMIGA), specifically designed by
\citet{MSS13} to deal with the interconnected process of galaxy
formation and reionization. In the present paper we show that,
combining the above mentioned information on reionization with the
current data on the evolution of the Universe at $z> 2$, we can
constrain, indeed, the freedom of this model, and have an
unprecedented detailed picture of the EoR.

The paper is organized as follows. In \sec \ref{amiga} we recall the
basic characteristics of AMIGA, paying special attention to its free
parameters. The observational data used to adjust those parameters are
given in \sec \ref{observables}, and the fitting strategy used is
described in \sec \ref{fitting}. Our results are presented in \sec
\ref{res}, and discussed and summarized in \sec \ref{conclude}.

Throughout the paper we assume the $\Lambda$CDM cosmology, with
$\Omega_\Lambda=0.684$, $\Omega_{\rm m}=0.316$, $\Omega_{\rm
  b}=0.049$, $h = 0.673$, $n_{\rm s}=0.965$, and $\sigma_8 = 0.831$
\citep{P15}.

\section{THE MODEL}\label{amiga}

AMIGA is a very complete and detailed, self-consistent model
of galaxy formation particularly well-suited to monitor the coupled
evolution of luminous sources and IGM. All processes entering the
problem are treated as accurately as possible according to our present
knowledge. In particular, AMIGA is the only model of galaxy formation
developed so far that includes \pIII stars, normal galaxies, and AGN,
together with an inhomogeneous multiphase IGM.

AMIGA deals with the properties of halos and their baryon content by
interpolating them in grids of masses and redshifts that are
progressively built, as halos merge and accrete, from trivial boundary
conditions. This procedure is less computer memory-demanding than the
usual method based on Monte-Carlo or $N$-body realizations of halo
merger trees, and enables us to reach very high redshifts and very low
masses while maintaining a good sampling over the entire range in
redshift and mass, respectively. In this way we can readily integrate
at every $z$ the feedback of luminous sources for the halo mass
function (MF), and accurately evolve the IGM.

The total number of free parameters in AMIGA is notably reduced in
comparison to ordinary SAMs thanks to causally linking physical
processes that are usually treated as independent from each
other. This greatly diminishes the degrees of freedom and renders the
model predictions more reliable. The reader is referred to
\citealt{MSS13} for a full description of AMIGA. Here we provide a
brief overview focusing on the model parameters.

\subsection{IGM}\label{main}

Ionizing photons have a very short mean free path in a neutral
hydrogenic gas (hereafter the gas outside halos is referred to as the
``IGM''). For this reason, luminous sources ionize (and reheat) the IGM in
small bubbles around them, which progressively grow and
percolate. \HII bubbles in turn harbor hotter \HeIII sub-bubbles due
to the most energetic fraction of UV photons emitted. Should the
ionizing flux become less intense, the singly or doubly ionized
bubbles stop stretching, and the surrounding \HI or \HeII regions
begin to grow again by recombination (or new dense \HI or \HeII nodes
begin to develop after complete reionization). In other words, the
neutral, singly, and doubly ionized IGM phases remain well-separated
except for small residual fractions.

X-ray photons produced in supernovae (SNe) and at a lesser extent in
\pIII stars, ordinary stars in binary systems, and AGN have instead a
large mean free path, and form a uniform background that reheats the
IGM by Compton scattering.\footnote{Compton scattering by CMB photons
  also heats the neutral phase until $z\sim 150$, when recombination
  is no more efficient, and the ionized fraction freezes out.} On the
contrary, adiabatic cooling due to cosmic expansion, collisional
cooling in hot neutral regions, and Compton cooling by CMB photons
oppose themselves to the heating. The latter mechanism is particularly
effective at high-$z$ where CMB photons are more abundant.

The intergalactic gas falling into halos is shock-heated to their
virial temperature. Thus the only halos able to trap gas at any given
moment are those with virial temperatures higher than the time-varying
temperature of the surrounding IGM. The accretion of gas onto galaxies
is calculated in AMIGA as usual in SAMs: taking into account the
cooling of the hot intrahalo gas and its free-fall to the halo
center. (Hereafter the gas trapped within halos is simply referred to
as ``the hot gas'', and the gas inside galaxies as ``the cold gas''.)
However, such a classical scenario strictly holds only in the case of
massive halos where the gas infall rate is limited by the cooling
rate. In low-mass halos it is limited by the rate at which the halo
accretes gas from outside. Since such an accretion is calculated in
AMIGA independently of the symmetry of the problem, our treatment thus
implicitly deals with accretion of cold gas onto galaxies according to
the filament scenario \citep{D09}.

Not only do galaxies accrete gas, but they also lose it through SN-
and AGN-driven winds. Thus the hot gas continuously incorporates
metals through three channels: i) via halo mergers, through the gas
within the merging halos; ii) via accretion, through the gas (if any)
in small accreted halos and directly in the form of diffuse gas
(inflows); and iii) via mass losses from the galaxies in the own halo
(outflows). Halos themselves may lose the hot gas if it becomes
unbound. AMIGA accurately takes into account all these mass 
exchanges between the different gaseous phases within ionized regions.

In neutral regions IGM remains unpolluted except for possible
recombination periods after complete reionization. Therefore, \pIII
stars with metallicity below some threshold value $Z\crit$ for atomic
cooling, in the range between $10^{-5}$ and $10^{-3}$ solar units, can
only form through \HHp-molecular cooling in such neutral, pristine
regions, which are immediately photo-dissociated and photo-ionized
around those stars.

The evolving extent of the different IGM phases is described by
means of the volume filling factors of singly and doubly ionized
regions, $Q\nHII\equiv \lav n\nHII\rav/\lav n\H\rav $ and
$Q\nHeIII\equiv \lav n\nHeIII\rav/\lav n\He\rav$, which result from
solving the improved differential equations \citep{MSS12}
\beq \dot Q\Si=\frac{\dot N\Si}{\lav n_{\rm S}\rav}
-\left[\!\left\lav  \frac{\alpha_{\rm
      SI}(T)}{\mu\se}\!\right\rav\bSi \!\frac{C\,\lav n\rav
  }{a^3}+\frac{\der \ln \lav n_{\rm S}\rav}{\der
    t}\right]Q\Si
\label{rec}
\eeq 
where a dot means time-derivative, with trivial initial conditions in
the dark ages. In equation (\ref{rec}) $t$ is the cosmic time,
subscripts S, SI, and SII mean either H, \HIp, and \HII or He, \HeIIp,
and \HeIIIp, angle brackets stand for the average over the region
indicated by the subscript (or over the whole IGM in the absence of
subscript), $n$ is the comoving IGM baryon density, $a$ is the cosmic
scale factor, and $\mu\se$ is the electronic contribution to the mean
molecular weight, i.e. $(\mu\se)\ii^{-1}$ is equal (neglecting metals) to
zero, $X+Y/4$, or $X+Y/2$, in neutral, singly, or doubly ionized
regions, respectively, $X$ and $Y$ being the usual hydrogen and helium
mass fractions.  $\dot N\Si$ is the comoving metagalactic ionizing
emissivity due to both luminous sources and recombinations, calculated
according to \citet{Me09},\footnote{For \HIp-ionizing photons, we
  include recombinations to \HeII and \HeI ground states. The
  contribution from \HeII \lya\ recombinations is neglected for
  simplicity.}  $\alpha_{\rm SI}$ is the temperature-dependent
recombination coefficient to the SI species, and $C\equiv \lav
n\nHII^2 \rav/\lav n\nHII\rav^2$, is the so-called clumping factor
giving the ratio between the real recombination factor and that in an
ideal homogeneous universe at a given $z$ (see below).

The term in claudators on the right of equation (\ref{rec})
takes into account inflows/outflows, and the volume average in region
SII of the function of temperature $f(T)\equiv \alpha_{\rm SI}(T)$ is
calculated using the Taylor expansion $f(T\ii)+(\der^2 f/\der
T\ii^2){\sigma_{\rm T\ii}^2}/2$ from the mean temperatures and
variances, $T\ii$ and ${\sigma_{\rm T\ii}^2}$, in phases i\ =\ I, II, and
III for neutral, singly, and doubly ionized regions, respectively,
solutions of the differential equations \citep{MSS12}
\beq
\frac{\der\ln T\ii}{\der\ln(1+z)}=
2+\frac{\der \ln (\mu\ii \varepsilon\ii/ n\ii)}{\der\ln(1+z)}\,
\label{Tigm}
\eeq
\beq 
\frac{\der \ln \sigma_{\rm T\ii}}{\der\ln(1+z)}= \frac{\der\ln T\ii}{\der\ln(1+z)}.
\label{T3}
\eeq
In equations (\ref{Tigm}) and (\ref{T3}) $\varepsilon\ii$, $n\ii$, and
$\mu\ii$ are respectively the phase-i comoving energy density,
comoving baryon density, and volume average mean molecular weight,
i.e.  $\mu\I^{-1}=X+Y/4$, $\mu_{\rm II}^{-1}=\mu\I^{-1} (1+Q\nHII)$,
and $\mu\III^{-1}=\mu\II^{-1} + Y/(4X)\, Q\nHeIII$ (neglecting
metals). The first term equal to 2 on the right of equation
(\ref{Tigm}) arises from adiabatic cooling, while the second term
includes the effects of Compton heating/cooling by CMB, Compton
heating by X-rays, heating/cooling by ionization/recombination of the
various hydrogen and helium species, cooling by collisional ionization
and excitation, achievement of energy equipartition of newly
ionized/recombined material, and inflows/outflows from
halos.\footnote{Neglecting non-linear effects (IGM is in linear and
  mildly non-linear regimes), the volume average of the
  heating/cooling by gravitational compression/expansion of density
  fluctuations vanishes.}

Thus the evolution of luminous sources determines through equations
(\ref{rec})--(\ref{T3}) the evolution of the IGM, and conversely, the
evolution of the IGM determines in the way shown next the evolution of
luminous sources, as well as the halo MF in regions with different
ionization state \citep{MSS12}. Once the evolution of all these
  components has been determined we can calculate the CMB optical depth
\beqa 
\tau\es = \Sigma_{\rm i=II}^{\rm
  III} \int_0^\infty\sigma\T\, \frac{c (1+z)^2\der z}{\mu\se\ii
  H(z)}\nonumber~~~~~~~~~~~~~~~~~~~~\\ \times\left[n\ii(z)+\int_{M\minii(z)}^\infty
  \!\!\der M \frac{M\hg(M,z)}{m\p} N\ii(M,z)\right],
\label{tauio}
\eeqa
and the Compton distortion $y$-parameter
\beqa 
y = \Sigma_{\rm i=II}^{\rm III}
\Bigg[\int_0^\infty\!\!\sigma\T\, \frac{c (1+z)^2\der z}{\mu\se\ii
    H(z)}\Bigg[n\ii(z)\frac{kT\ii(z)}{m\ee
      c^2}\nonumber~~~~\\+\int_{M\minii(z)}^\infty \!\!\der M\frac{M\hg(M,z)}{m\p} N\ii(M,z)\frac{kT\hg(M,z)}{m\ee
      c^2}\Bigg].
\label{y}
\eeqa 
In equations (\ref{tauio}) and (\ref{y}) $H(z)$ is the Hubble
parameter, $\sigma_{\rm T}$ is the photon Thomson scattering
cross-section, $c$ is the speed of light, $M\hg(M,z)$ and $T\hg(M,z)$
are the average mass and temperature of the hot gas in halos with mass
$M$, $N\ii(M,z)\equiv N_{\rm ion}(M,z)Q\ii(z)/(Q\II(z)+Q\III(z))$ is the
halo comoving abundance in region i, being $N_{\rm ion}(M,z)$ the halo MF in
ionized regions, and $M\minii(z)$ is the minimum halo mass in region i
able to trap gas, all of them dependent on $z$.

\subsection{Luminous Sources}\label{lum}

Halos grow through major mergers and smooth accretion. AMIGA follows
this growth in a well-contrasted analytic manner with no
free parameters \citep{RGS01,ssm98,Sea07}. This allows us to
accurately calculate, at any given moment, their abundance
\citep{JuEtal14a,JuEtal14b} and inner structure and kinematics
\citep{Sea12a,Sea12b}, which in turn sets the structure and
temperature of their hot gas \citep{Soea05}. The amount and
metallicity of such a hot gas and the properties of the central galaxy
and its satellites are monitored in AMIGA from the individual halo
aggregation history. 

As long as the hot gas has a metallicity below $Z\crit$, molecular
cooling takes place (see \citealt{MSS13}). When the temperature
reaches a minimum value that cannot be surpassed, the cold gas
accumulates in a central disk until the Bonnor-Ebert mass is
reached. The cloud then collapses and fragments, giving rise to a
small \pIII star cluster of about 1000 \modot.\footnote{The exact
  value depends slightly on $z$ as the temperature of the gas cloud is
  bounded by the CMB temperature.}  The initial mass function (IMF) of
\pIII stars is poorly known, but at the present stage we do not need
it. We only need the mass fractions $f_2$ and $f_3$ of \pIII stars
with masses in the ranges $130 \le M_\star/$\modot\ $ < 260$, and
$M_\star < 260$ \modot, respectively. Indeed, stars in the former
range explode in pair-instability SN (PISN), release about half their
mass in metals, and do not leave any black hole as remnant, while
those in the most massive range essentially collapse into a black hole
without producing metals \citep{HW02}. Thus, the yield of massive
\pIII stars, $p\III$, and the mass fraction of a \pIII star cluster
ending up in a coalesced mini-MBH, $\beta\III$, satisfy the
relations\footnote{If $\beta\III$ times the mass of the star cluster
  is less than 260 \modot, whether or not there is a mini-MBH of this
  latter mass is decided randomly using the \pIII star IMF (see
  Sec.~\ref{fitting}).}
\begin{equation}
p\III=0.5\times f_2\,
\label{y3}
\end{equation}
\beq 
\beta\III=1.0\times f_3\,.
\label{mu}
\eeq 
On the other hand, the \HIp- and \HeIIp-ionizing photon emissivities
of massive \pIII stars depend on the mass fraction $f_2+f_3$ in a
well-known way \citep{Schaerer02}.\footnote{\citet{Schaerer02}
  presumes the \pIII star IMF with the usual Salpeter slope. But the
  results are little sensitive to this assumption because the
  radiation emitted by all massive \pIII stars has essentially the
  same spectral energy distribution (SED), equal to that of a black
  body of $\sim 10^5$ K.} Lastly, \pIII stars with masses below 130
\modot\ contributing to the mass fraction $f_1=1-f_2-f_3$ behave like
ordinary \pIaII stars, except for their low metallicity. Therefore,
the feedback of \pIII stars is fully determined by the mass fractions
$f_2$ and $f_3$.

When the metallicity of the hot gas exceeds $Z\crit$, it undergoes
atomic cooling. The gas then contracts, keeping the initial angular
momentum, and settles in a rotationally supported disk. The structure
of disks is calculated according to \citet{MMW98} from the specific
angular momentum of the gas at the cooling radius, which equals that
of dark matter as provided by numerical simulations. This leads to
their effective (half-mass) radii $r\D$ and central surface density
$\Sigma(0)=M\D/(2\pi\,0.3\,r\D^2)$, where $M\D$ is the disk
mass.\footnote{Caution with the notation used in the present paper for
  the effective radius of a galaxy component. It coincides with that
  used in \citep{MSS12} for its length scale.}

If the disk is unstable, the cold gas is collected in a spheroid.
Spheroids also form in galaxy mergers. Indeed, when a halo is captured
by another one, it is truncated around the central galaxy which
becomes a satellite orbiting within the new halo. Satellites undergo
dynamical friction and orbital decay, being eventually captured by the
more massive central galaxy or merging with it if the mass ratio
between the two objects is greater than 1:3. Satellites also interact
between themselves and with the hot intrahalo gas. However, except for
ram-pressure stripping which is self-consistently calculated in
AMIGA, these interactions (tidal stripping, gas starvation at the
shock front of the hot gas, tidally induced disk-to-bulge mass
transfer,...) are neglected in the present work as they should play no
significant role at high-$z$ when groups and clusters of galaxies are
little developed and populated (but see Sec.~5.4).

Spheroids resulting from unstable disks or galaxy mergers have the
\citet{Hern} profile whose projection fits the $r^{1/4}$
law. Collisions between gas clouds yield the dissipative contraction
of recently assembled spheroids with mass $M\B$ as stars form,
according to the physically motivated differential equation for the
effective radius $r\B$ \citep{MSS13}
\beq 
r\B^2\,\dot r\B= 
-\frac{Z\cgb^{1/2}M\cgb} {Z_\odot^{1/2}\rho\dis\tilde\tau\acc},
\label{radius2}
\eeq 
from the initial value given by the $r\B-M\B$ relation of
non-contracted spheroids (see below) until the gas is exhausted or it
reaches the density of molecular clouds ($\sim 10^6$ particles
cm$^{3}$). In equation (\ref{radius2}) $M\cgb(t)$ and $Z\cgb(t)$ are
the mass and metallicity, respectively, of the cold gas in the
spheroid at $t$, $\tilde\tau\acc$ is the time elapsed from the
formation of the spheroid to the quenching of star formation by the
enlightened central AGN (see below), and $\rho\dis$ is a
characteristic gas density for dissipative contraction of spheroids.

When the cold metal-rich gas falls into a galactic component C, i.e. a
bulge (C=B) or a disk (C=D), new \pIaII stars form. The star
formation rate (SFR) satisfies the usual Schmidt-Kennicutt law
\begin{equation}
\dot M\sfc=\alpha\G\,{M\cgc\over\tau\dynX}\,,
\label{starf}
\end{equation}
where $M\cgc(t)$ is the mass of cold gas available, ${\tau\dynX}$ is
the dynamical timescale at the half-mass radius of the galactic
component, and $\alpha\G$ is the star formation efficiency.

The amount of interstellar gas heated by type II SNe in \pIaII star
formation episodes in the galactic component C is
\begin{equation}
M\rhc=\epsilon\C\, \frac{2\,\eta\SN E\SN}{V_{\rm C}^2-V\hg^2} M\sfc\,,
\label{m:reh}
\end{equation}
where $V_{\rm C}$ is the circular velocity at $r\C$, $V\hg$ is the
typical thermal velocity of the hot gas in the halo where the heated
gas is deposited, $E\SN=10^{51}$ erg is the typical energy liberated
by one typical SN explosion, $\eta\SN=0.014$ \modot$^{-1}$ is the
number of such explosions per unit stellar mass in a typical $\sim
0.20$ Gyr duration starburst for the adopted IMF, and $\epsilon\C$ is
the SN heating efficiency of the component. Due to the different SFR,
geometry and porosity of the gas in spheroids and disks, $\epsilon\C$
is let to take different values in bulges ($\epsilon\B$) and disks
($\epsilon\D$). Disks smaller than the corresponding spheroids are
however assumed to be oblate pseudo-bulges with $\epsilon\B$.

The amount of intrahalo gas heated through PISN during the explosion of
very massive \pIII stars and leaving the halo is also calculated
according to equation (\ref{m:reh}), but with $\epsilon\C=\epsilon\B$,
$E\SN=E\PISN=10^{53}$ erg, $\eta\SN=\eta\PISN=0.0015$ \modot$^{-1}$
\citep{HW02,Schaerer02}, $V\C$ equal to the halo circular velocity at
half-mass radius, and $V\hg=0$.

The instantaneous emission at all the relevant wavelengths of normal
galaxies is calculated according to their individual stellar formation
and metallicity histories using the stellar population synthesis model
(SPSM) by \citet{BC03} for the adopted IMF (see below) and taking into
account the typical yield of \pIaII stars \citep{Vea16} and all the
mass and metallicity exchanges between the different galaxy and halo
components.

In particular, as long as halos accrete, the metal-enriched gas
ejected from the central galaxy remains, due to viscosity, at the
cooling radius where it is the next to cool. AMIGA assumes that, at
the next major merger, all the hot gas of the progenitors is mixed up,
except for their inner $h\rec$ mass fraction which retains all the
re-heated gas remaining in those progenitors. This means that metals
ejected from central galaxies take at least two consecutive major
mergers to get well mixed within the halo. As a consequence, the hot
gas in the the inner part of the halo is always somewhat more
metal-rich than in the outer part.

MBHs originate as remnants of the most massive \pIII stars that, as
mentioned before, are supposed to coalesce in one mini-MBH per star
cluster. When \pIII star clusters (or their remnants) are subsequently
captured by normal galaxies and reach their spheroids, the
corresponding mini-MBHs migrate by dynamical friction to their centers
where they merge with the MBH lying there. MBHs at the center of
spheroids progressively grow not only through mergers, but mainly
through accretion of gas (and to a much lesser extent of stars). {\it
  AMIGA does not assume any specific fraction of the gas fallen into
  spheroids to fuel the central MBH}. Instead, it assumes that, each
time new gas falls into a spheroid, part of it forms stars, and part
fuels the MBH. The starburst lasts until it is quenched by the
enlightened AGN that heats mechanically an amount of gas equal to
\begin{equation}
M\rhb^{\rm AGN}=\epsilon\AGN \frac{2\,\langle L\rangle \tau\acc}{V_{\rm B}^2-V\hg^2}\,,
\label{m:rehAGN}
\end{equation}
and expels it back into the halo through super-winds. In equation
(\ref{m:rehAGN}) $\epsilon\AGN$ is the so-called quasar-mode AGN
heating efficiency, and $\lav L\rav$ is the AGN bolometric luminosity
$L(t)$, averaged over the typical duration ($\tau\acc= 0.15$ Gyr;
\citealt{Kellea10}) of one such enlightening episodes. The bolometric
light curve is modeled according to \citet{Hea03} through the
expression
\begin{equation}
L(t)=\dot M\BH V\last^2/2\,,
\label{lum2}
\end{equation}
where $V\last$ is the Keplerian velocity at the last marginally stable
orbit (at 9.2 times the Schwarschild radius) around the MBH, and $\dot
M\BH$ is the time-derivative of the MBH accretion curve, $M\BH(t)$,
assumed with a universal dimensionless bell-shaped form, whose maximum
at $\tau\acc$ sets the quenching of the starburst going on in the
spheroid.

The fraction of \HIp- (and \HeIIp-) ionizing photons escaping from
halos with a virial temperature below $10^4$ K is self-consistently
calculated by subtracting those photons ionizing the neutral intrahalo
gas, taking into account recombinations. Above $10^4$ K, a fixed
value, $f\esc$, is adopted, distinguishing between galaxies, $f\escG$,
and AGN, $f\escAGN$.

The fraction of the SN explosion energy converted to X-ray photons
through free-free emission of the SN remnant and inverse Compton
scattering of CMB photons by relativistic electrons is about 1\%
\citep{OH03}, and the fraction of the AGN bolometric luminosity
emitted in soft X-rays is 4\% \citep{VF07}. About half of those soft
X-ray photons Compton heat the IGM in neutral regions due to the
residual free electrons. The other half produce secondary ionizations
and excitations \citep{OH03}, neglected in AMIGA.

\subsection{Parameters}\label{param}

All quantities entering the previous equations are self-consistently
evolved in AMIGA from trivial initial conditions. Only a few of them
must be supplied as external inputs.

Some of those external inputs correspond to reasonably well known
functions of redshift, stellar mass, or galaxy mass. These are:

- The clumping factor: Hydrodynamic simulations show that $C(z)$ is
equal to about $3$ at $z\sim 6$ (\citealt{F12} and references
therein), and evolves essentially as
\beq
C_{\rm F}(z)=9.25-7.21 \log(1+z)\,.
\label{Fin}
\eeq
Strictly speaking, this fitting expression was derived by \citet{F12}
from their simulations where reionization started at $z\sim 13$, and
ended up at $z\ion\sim 8$. In the real Universe, $z\ion$ is instead
closer to $z\sim 6$, and starts at a redshift $z\ini$ to be
determined. We thus adopt the following generalization of expression
(\ref{Fin}),
\beq C(z)=C_{\rm F}(8)-\frac{[C_{\rm F}(8)-C_{\rm
      F}(13)]}{\log[(1+z\ini)/7]}\log[(1+z)/7]\,,
\label{Fin1}
\eeq
with identical linear behavior but satisfying $C(z\ini)\equiv C_{\rm
  F}(13)=1$ and $C(6)\equiv C_{\rm F}(8)=3$.

- The IMF of ordinary \pIaII stars: Observations show that it can be
approximated by a Salpeter IMF. In AMIGA we adopt the Salpeter slope,
$-2.35$, for large masses up to 130 \modot, and the slope $-1$ for
small masses in the range $0.1 < M_\star/\modotf <0.5$. Such an IMF is
consistent with the observed local one \citep{WTH08}, and similarly
top-heavy as the \citet{Cha03} IMF.

- The $r\B-M\B$ relation for non dissipatively contracted spheroids:
Assuming for simplicity that it is a power-law, $r\B=A M\B^{\gamma}$,
and taking into account that spheroids at the two mass ends of the
observed relations for nearby objects \citep{ShEtal03,ShEtal07,Phi03}
have not suffered dissipative contraction,\footnote{The most massive
  ellipticals have formed through dry major mergers, and spheroids
  with low enough mass reach the maximum density of molecular clouds.}
one is led to $A\approx 0.049$ Kpc \modot$^{-\gamma}$, and
$\gamma\approx 0.20$.

Even though there is some uncertainty in the preceding expressions,
the results of the present study have been checked to be very robust
against reasonable variations in the coefficients.

But most of the external inputs are constant or can be considered as
such in a first approximation.\footnote{For instance, the escape
  fractions from galaxies and AGN could vary with $z$. However,
  neither from a theoretical nor an observational point of view is
  that dependence clear. Thus we assume them with fixed typical
  values, as usual.}  Many of them are either well determined, as in
the case of the threshold mass ratio for galaxy captures with
destruction, the yield of \pIaII stars, and the energy and frequency
of SNe, or they play an insignificant role in the results, as the
fraction of the energies released by SNe and AGN that are converted
into X-rays, so they can safely be taken with the fixed values
specified in Section 2.2. Others can also be taken with any reasonable
fixed value because their uncertainty is absorbed in other
parameters. This is the case of the factors 0.5 and 1.0 in equations
(\ref{y3})--(\ref{mu}), the constant $\tilde \tau\acc$ in equation
(\ref{radius2}), and the product $V\last^2\tau\acc$ times the factor
arising from the universal dimensionless MBH accretion curve in
equation (\ref{m:rehAGN}).

However, all the remaining external inputs must be left free since
they are poorly determined, and have a significant effect on the
results. These are thus the real parameters of the model to be
adjusted through the fit to the observational data. Their complete
list is the following:

\indent $Z\crit$: threshold metallicity for atomic cooling,\\ \indent
$f_2$: mass fraction in intermediate \pIII stars,\\ \indent $f_3$:
mass fraction in massive \pIII stars,\\ \indent $\rho\dis$:
characteristic density for dissipative contraction,\\ \indent
$\alpha\G$: Pop I \& II star formation efficiency,\\ \indent
$\epsilon\B$: SN heating efficiency in spheroids, \\ \indent
$\epsilon\D$: SN heating efficiency in disks, \\ \indent
$\epsilon\AGN$: AGN quasar mode heating efficiency, \\ \indent
$h\rec$: mixing hot gas mass fraction, \\ \indent $f\escG$: escape
fraction from galaxies, and \\ \indent $f\escAGN$: escape fraction
from AGN.

\noindent As we will see in Section \ref{fitting}, there is some
degeneracy between these 11 parameters. In addition, the unknown \pIII
star IMF must satisfy some conditions that have not yet been
enforced. As a consequence we will end up with only 9 degrees of
freedom. Although this may still seem quite a large number, it is
similar to that found in all recent analytic models of the
EoR\footnote{In all of them the clumping factor and the IMF of
  ordinary stars are also fixed, as well as the uncertain
  $z$-dependent dust attenuation.}  that find $z\ion=6$ from the fit
to the observed SFR density of galaxies at $z>6$
\citep{Rea15,I15,BouwEtal15}. We remark, however, that our model is
expected to recover not only the observed redshift of full hydrogen
ionization, but the whole evolution of the Universe from the fit to
all the observed cosmic histories at $z>2$. Such an achievement with
so few parameters is possible thanks to the fact that our model is
{\it self-consistent} so that many quantities can be calculated and
need not to be considered as free parameters.

\section{The Observed high-z Universe}\label{observables}

\begin{figure*}
\centering
 {\includegraphics[scale=1.,bb=198 138 365 726]{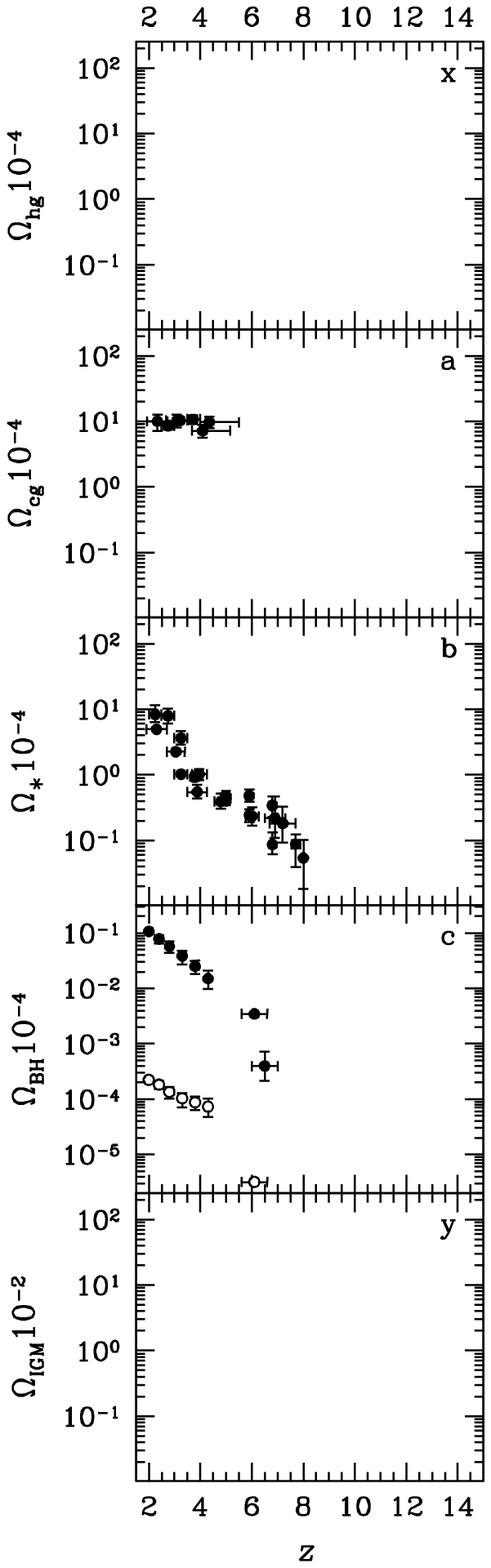}}
 {\includegraphics[scale=1.,bb=196 138 365 726]{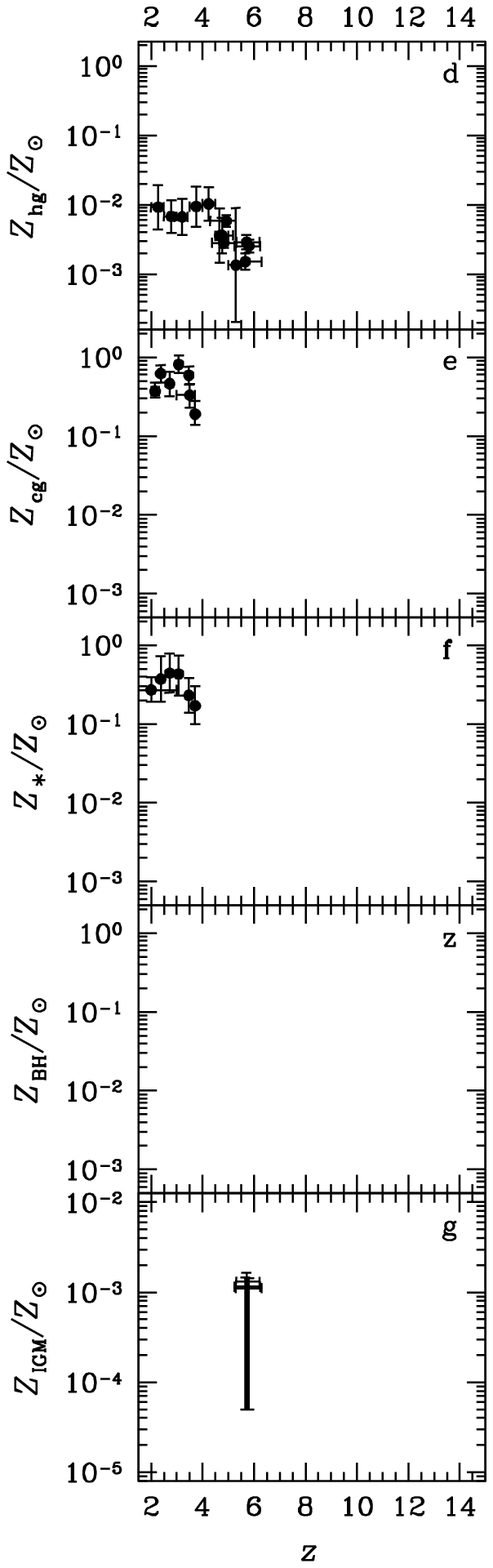}}
 {\includegraphics[scale=1.,bb=196 138 365 726]{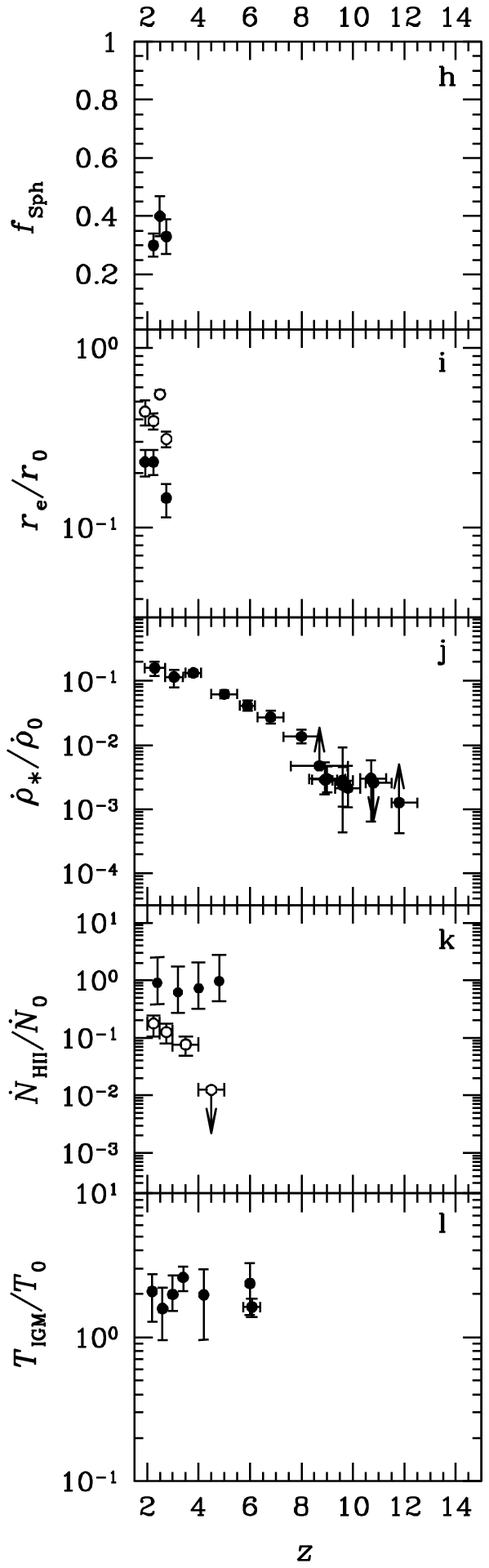}}
\vspace{-20pt}
 \caption{Cosmic histories with available data at $z> 2$ (see text
   for the different symbols in some panels): (a) cold gas mass
   densities \citep{PMSLI03,PHFW05}; (b) stellar mass densities
   \citep{RS09,StarkEtal09,GonzEtal10,LabEtal10a,LabEtal10b,GonzEtal11,MortEtal11a,Cap11};
   (c) MBH mass densities \citep{Kellea10,WillEtal10b,Trea11}; (d) hot
   gas metallicities \citep{Songaila01,RPMZ09,SimEtal11,OEtal13}; (e)
   cold gas metallicities \citep{Ma08,So12,Cu14}; (f) stellar
   metallicities \citep{Ha08,So12}; (g) IGM metallicities (based on
   estimates by \citealt{RPMZ09,SimEtal11,OEtal13}); (h) fraction of
   spheroid-dominated galaxies with masses $\ga 10^{11}$
   \modot\ \citep{Betal08,BEtal14a}; (i) median effective radii of
   spheroid-dominated galaxies \citep{Betal08}; (j) SFR
   densities
   \citep{RS09,LorEtal11,BouwEtal12b,OeschEtal13,CoeEtal13,EllisEtal13};
   (k) total ionizing emissivities \citep{BB13} and the contribution
   from AGN \citep{CEtal09}; and (l) IGM temperatures
   \citep{LiEtal10,BBWHS10,Bolea12}. Error bars are 1$\sigma$
   deviations, except for panel (g) where they give upper and lower
   limits. The density parameters are for the {\it current value} of
   the critical cosmic density, $\rho\crit(z=0)$, and the scaling
   factors in the right column are: $r_0=r\ee(z=0)$, $\dot\rho_0=1$
   \modot\ yr$^{-1}$ Mpc$^{-3}$, $\dot N_0=10^{51}$ photons s$^{-1}$
   Mpc$^{-3}$, and $T_0=10^4$ K. Other cosmic histories with no data
   also included for the densities and metallicities in the left and
   middle columns to cover all baryonic components: (x) hot gas mass
   densities, (y) IGM mass densities, and (z) metallicities of matter
   fallen into MBHs.}
\label{data}
\end{figure*}

All the observational data currently available on the evolution of the
Universe at $z> 2$ (see Fig.~\ref{data}) refer to: \\ \indent (a)
the cold gas mass density history (CGH), \\ \indent (b) the stellar
mass density history (STH), \\ \indent (c) the MBH mass density
history (MBHH), \\ \indent (d) the hot gas metallicity history (HGMH),
\\ \indent (e) the cold gas metallicity history (CGMH),\\ \indent (f)
the stellar metallicity history (STMH), \\ \indent (g) the IGM
metallicity history (IGMMH), \\ \indent (h) the galaxy morphology
history (GAMH), \\ \indent (i) the galaxy size history (GASH),
\\ \indent (j) the SFR density history (SFH), \\ \indent (k) the
\HIp-ionizing emissivity history (IEH), and\\ \indent (l) the IGM
temperature history (IGMTH).\\ Taking into account that there are two
independent IEHs, one for galaxies and the other for AGN, this
represents 13 independent global, i.e. averaged or integrated, cosmic
histories.\footnote{The STH is the time-integral of the SFH, but the
  two observables are inferred independently. In fact, as we will see,
  they turn out to be inconsistent with each other.}  In addition, the
following differential properties are also available at a few discrete
redshifts (see e.g. Figs.~\ref{Gal_MFs} and \ref{MBH_MFs}): \\ \indent
(m) the galaxy stellar MFs (or UV LFs), and \\ \indent (n) the MBH MFs
(or AGN optical and X-ray LFs).\\ The mass density and emissivity
estimates at different redshifts are derived from these MFs or LFs at
the corresponding $z$, extrapolated beyond the observational
limits.\footnote{Given the asymptotic logarithmic slopes of those MFs
  or LFs, such extrapolations are not expected to have any dramatic
  effect (but see Sec.~5.3).} Therefore, these MFs or LFs harbor more
  information than the corresponding global properties above. However,
  they are harder to handle. Thus we will concentrate in trying to fit
  the former cosmic histories, and then check whether the galaxy and
  MBH MFs at a few representative redshifts are also well recovered.

We briefly discuss next the techniques, approximations, and models
used to infer all those data. All quantities given below refer to the
Salpeter-like IMF adopted in AMIGA (applying the conversion factor
from a strict Salpeter IMF provided by \citealt{WTH08}) and to a
Hubble constant of 67.3 km s$^{-1}$ Mpc$^{-1}$.

(a) Cold gas mass densities are computed from the distribution function
of \HI column densities per unit redshift in damped \lya\ systems
(DLAs) in the spectra of quasars up to $z\ion$, integrated from zero
to infinity. The estimates shown in panel (a) of Figure \ref{data}
were obtained by \citet{PMSLI03} and \citet{PHFW05}. Note that some
amount of \HH (and other molecules), not accounted for in these
estimates, must also be present. However, the real gas densities are not
expected to be substantially greater.

(b) Stellar mass densities are inferred by fitting the SED of high-$z$
galaxies, obtained from several optical and infrared rest-frame bands,
to template SEDs generated by SPSMs assuming an IMF and a dust
absorption correction.\footnote{Other quantities, such as the
  metallicity or the time elapsed since the last starburst, are
  usually varied within reasonable ranges and typical (or favorite)
  values are adopted.} The resulting galaxy stellar MFs are
extrapolated down to $10^8$ \modot. In panel (b) we show one
illustrative sample of datasets obtained by several authors
\citep{RS09,StarkEtal09,GonzEtal10,LabEtal10a,LabEtal10b,GonzEtal11,MortEtal11a,Cap11}. The
large scatter in the data is mostly due to the very uncertain
correction for dust attenuation.

(c) MBH masses at high redshifts are estimated by the method of
reverberation mapping that assumes the broad line emitting clouds
around AGN are virialized. The largest uncertainty in those masses
arises from the unknown geometry and inclination of the system, and is
encapsulated in a virial factor ($f_\sigma$ or $f_{\rm FWHM}$,
depending on whether the potential well is traced through the cloud
velocity dispersion or the emission-line width, respectively) of order
one. The MBH mass density estimates at $z< 5$ and $z=6.1$ plotted as
open circles in panel (c) have been computed using the MBH MFs derived
by \citet{Kellea10} and \citet{WillEtal10b}, respectively, from AGN
LFs in rest-frame UV. These MFs are, however, greatly affected by AGN
obscuration and inactivity, and are substantially lower than the
filled circle at $z=6.5$ inferred by \citet{Trea11} from the
extrapolated AGN LF in X-rays, less affected by these
effects.\footnote{Only about 20-30 \% of all AGN are believed to be
  Compton thick, and hence lost in X-rays \citep{CM13}. On the other
  hand, most of them are active at $z\ga 6$ \citep{WillEtal10b} due to
  the high frequency of unstable disks (see Sec.~\ref{res}).}  On the
contrary, this latter estimate is fairly aligned with the filled
circles at the previous redshifts, derived assuming a constant
MBH-to-spheroid mass ratio, $\mu$, equal to $8\times 10^{-3}$
(\citealt{KH13}, taking into account the change in
IMF),\footnote{Observations show that the MBH-to-spheroid mass ratio
  is constant at least up to $z\sim2$. The mild evolution seen beyond
  that redshift is likely due to the Malmquist bias \citep{KH13}.} and
the upper envelope of the stellar mass densities given in item
(b).\footnote{We adopt the upper envelope because it is more
  consistent with the SFR density estimates of item (j).}

(d) Absorption lines of metal ions, usually \CIVp, are observed outside
the \lya\ forest in the spectra of distant quasars. The absorbers have
column densities $N$ according to a distribution function at any given
$z$ well-fitted by the expression $f(N)\propto N^{-(1.32+0.0779 z)}$
\citep{OEtal13}, from which one can infer the average metal abundance
of the absorbers. The metallicities plotted in panel (d), obtained by
\citet{Songaila01}, \citet{RPMZ09}, \citet{SimEtal11}, and
\citet{OEtal13}, have been homogenized to the widest column density
range available ($10^{12}-10^{15}$ cm$^{-2}$), converted to C
abundances for a [\CIVp/C] abundance ratio of 0.05 appropriate to high
column densities \citep{Sim06}, and then to metallicities assuming the
C mass fraction of ejecta from ordinary Pop I \& II stars given by
\citep{RPMZ09}. Indeed, the absorbers are believed to be neutral
polluted gas regions around intervening halos, with density contrasts
spanning from 10 to 100 in redshifts going from $6$ to $2$,
respectively. Consequently, these metallicities should coincide with
that of the hot gas in low-mass halos having accreted gas for the
first time (i.e. with masses between the minimum trapping mass at that
$z$ and, say, 1 dex larger).

(e) Gas-phase metallicities of high-$z$ star-forming galaxies are
derived from emission lines in their composite spectra. The estimates
obtained for different metal ions are calibrated against each other,
and converted to the [O/H] abundance ratio, then to the global
metallicity, using the relation provided by \citet{So12}. In panel (e)
we plot typical values obtained for small samples of galaxies with
stellar masses above $\sim 5\times 10^{9}$ \modot\ at $z$ around $3.5$
and $2.2$, respectively inferred by \citet{Ma08} and \citet{Cu14}, and
several individual values for a few galaxies with similar masses at
the remaining redshifts inferred by \citet{So12}.

(f) Stellar metallicities of high-$z$ star-forming galaxies are derived
similarly to the gas-phase metallicities in item (e) but from ion
absorption lines. In panel (f) we plot the values obtained for a small
sample of galaxies with stellar masses above $\sim 5\times 10^{9}$
\modot~at $z$ around 2.0 inferred by \citet{Ha08} and several
individual values corresponding to a few galaxies with similar masses
inferred by \citet{So12}.

(g) There are no direct measures of the average IGM
metallicities. However, we can put in panel (g) some upper and lower
limits. Indeed, as mentioned in item (d), metal absorption lines
outside the \lya\ forest in the spectra of distant quasars would be
produced by the polluted IGM around halos. Since the metallicity of
those systems decreases with increasing $z$ due to the increasing
effect of galaxy ejecta, their values for the most distant systems
($5\le z < 6$) are the best upper limits to the average ``primordial''
(i.e. enriched by \pIII stars only) IGM metallicity we can have. For
the present purposes, these metallicities are derived assuming a
[\CIVp/C] abundance ratio of 0.5 \citep{RPMZ09}, appropriate to low
column densities, and a carbon mass fraction of \pIII stars (model C
in \citealt{Schaerer02}) as it corresponds to IGM polluted by \pIII
stars only. On the other hand, the IGM metallicity can become lower
than $Z\crit$ at low-$z$ due to the lower metallicity-to-ionizing photon
contribution of normal galaxies into the IGM compared to that of \pIII
stars that dominate at high-$z$. A lower limit of $0.5\times 10^{-4}$
Z$_\odot$ is thus an educated guess.

(h) Morphological fractions are inferred from fitting a \citet{Sersic}
law to galaxy surface-brightness profiles using circular
apertures. This yields the S\'ersic index, $n$, and the effective
(half-light) radius, $r\ee$, of the galaxy. Alternatively one can
decompose the surface-brightness profile in a bulge and a disk. The
fractions of spheroid-dominated star-forming galaxies with stellar
masses $\sim 10^{11}$ \modot\ plotted in panel (h) were obtained by
\citet{Betal08} for a threshold S\'ersic index value of $n> 2$, and by
\citet{BEtal14a}, at $z=2.5$, for star-forming galaxies with
bulge-to-total light ratios greater than 0.5. Both methods yield
similar results.

(i) Galaxy sizes depend on the galaxy morphology. The median effective
radii of spheroid-dominated ($n> 2$) and disk-dominated ($n\le 2$)
objects scaled to the corresponding local values \citep{ShEtal03} for
galaxies with stellar masses $\sim 10^{11}$ \modot\ plotted in panel
(i) as filled circles and open circles, respectively, were obtained by
\citet{Betal08} for the same galaxy sample as used to draw the
morphological fraction in item (h). The estimates inferred by
\citet{BEtal14b} directly using the effective radii of bulges and
disks yield substantially different results and have not been
considered. On the other hand, we will concentrate, from now on, in
the effective radii of spheroid-dominated galaxies, as those of
disk-dominated galaxies depend on the relative spatial distribution of
stars and gas in the disk which is not modeled in AMIGA.

(j) SFR densities, $\dot\rho_\ast$, are derived from the luminosity
of observed galaxies in broadband or narrowband filters that traces
active star formation, correcting for incompleteness from the
extrapolated LFs and for dust absorption, applying IMF-dependent SFR
calibrations. In panel (j) we plot the estimates inferred by
\citet{RS09}, \citet{LorEtal11}, \citet{BouwEtal12b},
\citet{OeschEtal13}, \citet{CoeEtal13}, and \citet{EllisEtal13}. The
dust-uncorrected values listed by \citet{BouwEtal12b} have been
shifted upwards according to \citet{Jaa12} to correct for that
effect. The resulting SFR densities are in very good agreement with
the latest results published by \citet{McLeod15}.

(k) Similarly to the escape fraction of ionizing (or Lyman-continuum,
\lyc) photons, which has two different versions, one for galaxies and
the other for AGN, we will distinguish between the \HIp-ionizing
emissivity from normal galaxies and AGN. At high-$z$, the
\lyc\ emissivity of normal galaxies is usually estimated from their
SFR densities assuming a given escape fraction of ionizing photons
\citep{BouwEtal11,RiEtal06,StarkEtal07}. These estimates are not
considered here as they are equivalent to the SFR densities plotted in
panel (h). An alternative method \citep{BB13}, {\it independent from
  the SFR densities} consists of using the \lya\ effective optical
depth inferred from the comparison of the observed \lya\ forest with
synthetic data obtained from hydrodynamic simulations, modeling photon
mean free paths for the assumed ionizing emissivities. These are the
estimates plotted as filled circles in panel (k). Regarding to
\lyc\ emissivity from AGN, this is estimated from UV observations of a
large sample of X-ray AGN. The results derived by \citet{CEtal09} are
plotted in the same panel as empty circles.

(l) The temperature of the singly ionized IGM is usually derived from
the small-scale structure of the \lya\ forest, using hydrodynamic
simulations and galaxy and quasar emission models. This is the case
for the data at $z<5$ derived by \citet{LiEtal10} plotted in panel
(l). We do not include the estimates inferred by \citet{BBHS11} as
they greatly depend on the poorly known density-temperature
relation. Unfortunately, all those measures are extremely challenging
for $z$ approaching 6. An alternative method that sidesteps such a
difficulty consists of probing the IGM temperature within a proper
distance of $\sim 5$ Mpc of quasars by combining the cumulative
probability distribution of Doppler broadening and synthetic
\lya\ spectra. The measurements plotted at $z \sim 6$ were obtained in
this way by \citet{BBWHS10} and \citet{Bolea12}, using one and 7
quasars, respectively.

(m) Galaxy stellar MFs are obtained from the galaxy LFs in rest-frame
UV. The conversion involves the assumptions and procedures mentioned
in item (b). For consistency, we adopt the MFs corrected for
absorption inferred by \citet{GonzEtal11} at $z=3.8$ and $z=6.8$,
which were also used to derive the stellar mass densities plotted in
panel (b) at those redshifts that typically bracket the range of
interest.

(n) Similar comments hold for the MBH MFs. Their derivation from
optical AGN LFs involves the assumptions and procedures mentioned in
item (c). For consistency, we adopt the MFs inferred by \citet{VeEtal}
and \citet{WillEtal10b} at $z=3.3$ and $z=6.1$, respectively, which
were also used to infer the MBH mass densities plotted in panel (c) at
those redshifts that bracket the range of interest. 

\section{Fitting strategy}\label{fitting}

The goal is to adjust the 11 free parameters mentioned in \sec
\ref{amiga} through the fit to the 13 independent cosmic histories
described in \sec \ref{observables}, enforcing the constraints on
$z\ion$ and $Q\nHII(z)$ mentioned in Section \ref{intro} as priors.

Scanning a whole 11-D parameter space is not an easy
task. Fortunately, we can take advantage of some particularities of
the problem that greatly simplify the fitting procedure.

First, the characteristic dissipation density, $\rho\dis$, is fully
constrained by the data, regardless of the value of any other
parameter. Indeed, equation (\ref{radius2}) recovers the observed
effective radius of massive spheroids at $z=2-3$ formed in major
mergers of two galaxies with the typical gas content and metallicity
(see panels (i), (a), and (e) of Fig.~\ref{data}) provided only
$\log(\rho\dis$ \modot$^{-1}$\ Kpc$^{3})=6.0\pm 0.6$.

Second, the properties of normal galaxies and MBHs decouple from those
of \pIII stars \citep{MSS13}. Indeed, massive \pIII stars lose metals
through PISN-driven winds into the ionized bubbles they form
around. When this polluted gas falls into halos and cools, normal
galaxies form. In principle, the properties of those galaxies should
thus depend on the metallicity of the ionized IGM set by \pIII stars
through the values of $Z\crit$, $f_2$, and $f_3$. But normal galaxies
eject such large amounts of metals into the hot gas in halos through
type II SN- and AGN-driven winds that the hot gas rapidly loses the
memory of its initial metallicity, and the properties of galaxies
rapidly become independent of $Z\crit$, $f_2$, and $f_3$. Similarly,
MBHs are seeded by the remnants of massive \pIII stars. But the
dramatic growth of MBHs within spheroids, regulated by the amount of
gas reaching them, causes MBHs to rapidly lose the memory of their
seeds, and the properties MBHs also become independent of $Z\crit$,
$f_2$, and $f_3$. We can thus concentrate in adjusting, in a first
step, the parameters referring to \pIII stars and, afterwards, those
referring to normal galaxies and MBHs. Notice that, even though the
properties of the two kinds of sources are independent of each other,
we cannot exchange the order of those fits. Some of the observables
involving the parameters of the latter set refer to mean mass
densities per unit volume, not per unit {\it ionized} volume, or to
metallicities averaged over all regions, not over {\it ionized}
regions. Consequently, they involve not only the properties of normal
galaxies and MBHs, but also the volume filling factor of ionized
regions, which depends on the mass fraction, $f_2+f_3$, of massive
\pIII stars. Nonetheless, once the parameters $f_2$ and $f_3$ are
fixed, those observables will depend only on the parameters referring
to normal galaxies and MBHs.

Third, most of the parameters in these two sets can be adjusted
sequentially. 

For a given value of $Z\crit$, the ionization and
photo-heating of the IGM at high-$z$ depend only on the mass fraction
$f_2+f_3$ of massive \pIII stars, while the mass of metals ejected in
ionized regions depends only on the mass fraction $f_2$. Thus it
should be possible to adjust $f_2+f_3$ by fitting the IGMTH at
high-$z$, and then adjust $f_2$ by fitting the IGMMH also at high-$z$.

On the other hand, for a given couple of $\epsilon\B$ and $\epsilon\D$
values, the amount of gas in the disk or spheroid of a galaxy at any
moment is equal to the amount of gas fallen into it, which depends on
cosmology (through the halo characteristics, the hot gas cooling rate,
and the galaxy merger rate),\footnote{Cooling is much more sensitive
  to the density of the hot gas than to its metallicity, dependent on
  $h\rec$.} minus the amount of gas used to form stars and ejected
from the component (the latter depending on the former).\footnote{The
  gas mass fueling the MBH is negligible compared to the mass
  converted into stars.} The inductive reasoning thus implies that the
stellar masses of normal galaxies depend only on
$\alpha\G$. Similarly, the mass of gas fueling the central MBH of a
galaxy depends on the gas mass collected in the spheroid, which
together with the stellar mass sets the dynamics, and hence, the mass
of newly formed stars and of gas heated by SNe, the mass of the MBH
determining its accretion rate, and the mass of gas heated by the
enlightened AGN. Therefore, the inductive reasoning implies that the
MBHH depends on both $\alpha\G$ and $\epsilon\AGN$. Lastly, the
metallicity of the hot gas in the halo that determines, through the
gas cooling and the metal-enriched gas ejected from galaxies, the cold
gas and stellar metallicities of those objects depends on cosmology,
on $\alpha\G$, $\epsilon\AGN$, and on $h\rec$ setting the mixing rate
of the reheated gas in the halo. Finally, the emissivity of galaxies
and AGN depends on all the previous parameters setting the whole
evolution of galaxies and AGN and their respective intrinsic ionizing
photon production rates and the values of $f\esc$ from the two kinds
of sources. Consequently, it should be possible to adjust $\alpha\G$
by fitting the SFH (or STH), then $\epsilon\AGN$ by fitting the MBHH,
then $h\rec$ by fitting the HGMH (or any of the CGMH and STMH), and
lastly the $f\esc$ values of galaxies and AGN by fitting their
corresponding IEHs with the suited priors on $z\ion$ and $\der_z
Q\nHII$ (hereafter $\der_z$ means $z$-derivative).

Of course, this sequential fitting procedure should be carried out for
every possible set of $Z\crit$, $\epsilon\B$, and $\epsilon\D$ values
in the corresponding 3-D space. But this is a minor difficulty
compared to directly looking for acceptable solutions in the whole
11-D parameter space.

Any ``acceptable solution'' should thus fit with acceptable $\chi^2$
values the IGMMH, the IGMTH, the SFH (and the STH), the MBHH, the HGMH
(and the CGMH and STMH), and the IEHs for galaxies and AGN, with
$z\ion=6.0^{+0.3}_{-0.5}$ and $\der_z Q\nHII(z=7)< 0$. Moreover, since
$\epsilon\B$ and $\epsilon\D$ govern the amount of cold gas remaining
in disks and of stars forming in spheroids, the CGH and GAMH should
hopefully also be fitted. And given the way $\rho\dis$ is determined,
the GASH should too. In principle, the solution so obtained should
also recover the observed galaxy stellar and MBH MFs and, provided
$Z\crit$ is not degenerate so that there is still one degree of
freedom to play with, we may expect the solution to also fulfill the
constraints on $\tau\es$, and $y$. Thus the problem is not only
well-posed, but in principle also slightly overdetermined.

Unfortunately, the IGMMH and IGMTH at high-$z$ are not
well-constrained. The only available data on the IGM temperature refer
to $z\la 6$ where \pIII stars play no significant role, and there are
just a few loose bounds for the IGM metallicity. Consequently, $f_2$
and $f_3$ cannot be determined in the previous simple
manner. Moreover, $Z\crit$ is degenerate with $f_2$ and $f_3$. The
good news is that this degeneracy reduces the degrees of freedom to
10, which facilitates the fit at the cost, of course, of making it
even harder to find any acceptable solution.

Indeed, massive \pIII stars are the first source of ionizing photons,
with emissivity equal to $m\p \xi\III(f_2+f_3)\dot M\III$, where $\dot
M\III$ and $\xi\III$ are the formation rate and \lyc\ photon
production rate of \pIII stars, respectively. Those stars also pollute
with metals the ionized bubbles at the rate $0.5f_2\dot
M\III$. Therefore, the formation of normal galaxies is possible
provided the metallicity of such bubbles,
$0.5f_2/[m\p\xi\III(f_2+f_3)]$, is greater than $Z\crit$. But this
condition is necessary although not sufficient. When normal galaxies
form, they keep on ionizing and photo-heating the surrounding IGM,
while they lose very few metals into it (they rather enrich the hot
intrahalo gas). Consequently, the metallicity of ionized bubbles
decreases, and their temperature increases, so the formation of new
galaxies may be quenched, and reionization may be aborted. Only if the
initial IGM metallicity is {\it sufficiently large} will the formation
of normal galaxies last long enough for reionization to proceed
successfully.

According to the previous discussion, the evolution of the Universe
for some values $Z\crit$, $f_2$, and $f_3$ will thus be essentially
the same (with $Z\IGM$ in units of $Z\crit$) as for other values
$Z\crit'$, $f_2'$, and $f_3'$, provided the IGM reionization and
metal-enrichment histories driven by \pIII stars are the same, that is
provided the two sets of values satisfy the relations
\beq 
f_2'+f_3'=f_2+f_3
\label{deg1}
\eeq
\beq 
\frac{f_2'}{(f_2'+f_3')Z\crit'}=\frac{f_2}{(f_2+f_3)Z\crit}\,.
\label{deg2}
\eeq

This leads to the following simple fitting procedure. For some
fiducial values of $Z\crit$, and $f_2/(f_2+f_3)$ warranting that
reionization will not be aborted, we scan all possible values of
$f_3$, $\epsilon\B$, and $\epsilon\D$, and for each set of these three
values we adjust $\alpha\G$, $\epsilon\AGN$, $h\rec$, and $f\esc$ in
the sequential manner explained above, with the priors on $z\ion$ and
$\der_z Q\nHII$. Then, given any acceptable solution, the relations
(\ref{deg1}) and (\ref{deg2}) provide with any other combination of
$Z\crit$, $f_2$, and $f_3$ values leading to it. Lastly, we can check
the values of $\tau$ and $y$ as well as the right behavior of the
galaxy stellar and MBH MFs.

\subsection{Additional Constraints on the Pop III Star IMF}

Were the \pIII star IMF known, we could readily calculate the mass
fractions $f_2$ and $f_3$, and, as a bonus, leave the above mentioned
degeneracy. Unfortunately, such an IMF is poorly determined. But we
can still enforce some conditions such an IMF must satisfy in order to
further constrain the problem and reduce the degrees of freedom to 9
(8 if we discount $\rho\dis$, directly determined by the data).

The IMF of \pIII stars must be top-heavier than the IMF of ordinary
stars \citep{L98}.\footnote{The rationale for a top-heavier \pIII IMF
  is that the fragmentation of protostellar clouds is favored by
  metals, absent in the pristine gas.} Approximating it by a power-law
$f(M_\star)\propto M_\star^{-\alpha\III}$ like the Salpeter IMF, this
implies either that $\alpha\III$ is smaller than the Salpeter slope,
2.35, or that the upper or lower stellar masses are larger than for
ordinary stars, i.e. $M\III\lo > 0.5$ \modot\ and $M\up\III> 130$
\modot,\footnote{The upper mass of Pop I \& II stars is poorly
  determined. An upper limit of 150 \modot\ is sometimes considered,
  although no stars more massive than 130 \modot\ have ever been
  observed (\citealt{F05}; but see \citealt{CEtal10}).} or both
conditions at the same time.

\begin{figure}
\centering
{\includegraphics[scale=.44,bb=5 14 700 590]{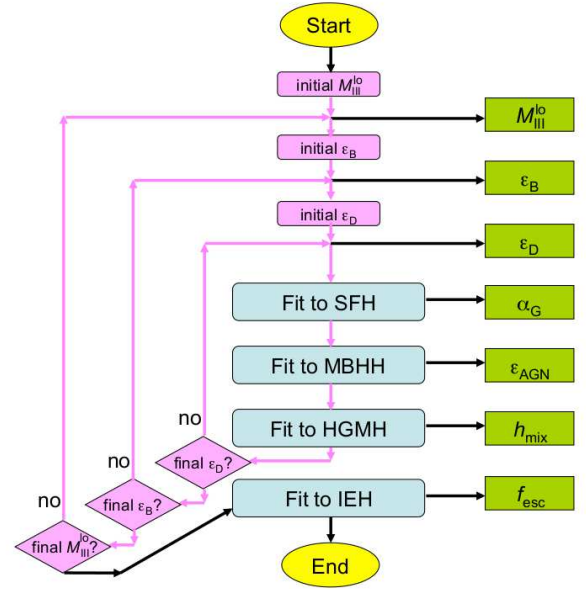}}
\caption{Fitting procedure making use of the \pIII star IMF and some
  fiducial values of $Z\crit$, $\alpha\III$, and $M\III\up$.}
\label{Flowchart_right}
\end{figure}

But this is not all. The fact that the slope of the IMF for ordinary
Pop I \& II stars appears to be little sensitive to their metallicity
indicates that $\alpha\III$ should not be substantially smaller than
$2.35$;\footnote{This is the reason why in most studies dealing with
  the \pIII star IMF the value of 2.35 is adopted
  (e.g. \citealt{Schaerer02}).} i.e. it should be larger than 2.25 or
at most than 2.15. On the other hand, $M\III\lo$ should be smaller
than 260 \modot, otherwise \pIII stars would not enrich with metals
the IGM, and $M\III\up$ should be larger than 260 \modot, otherwise
\pIII stars would not seed MBHs, and smaller than $500$ \modot\ or at
most than 1000 \modot\ \citep{HW02,Schaerer02}.

\begin{figure*}
\centering
{\includegraphics[scale=.7,bb=60 270 550 550]{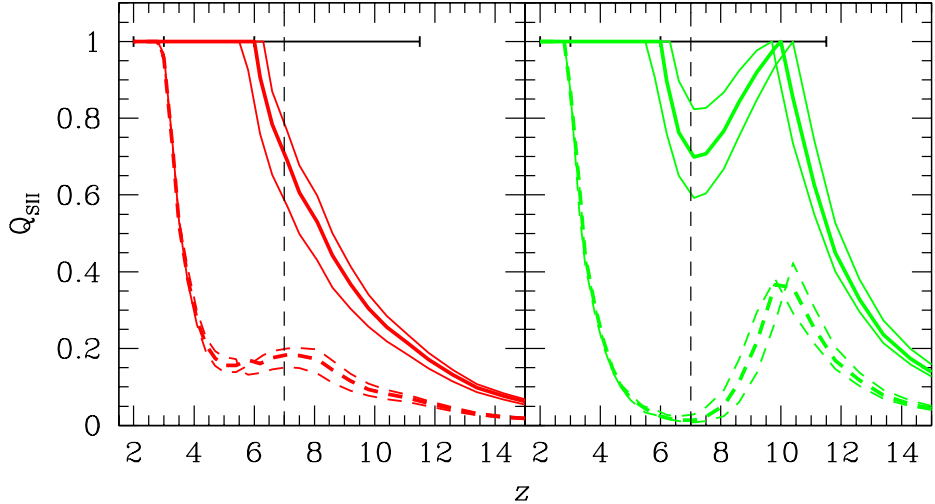}}
 \caption{Only acceptable solutions with single (left panel) and
   double (right panel) reionization. In thick lines models S and D
   giving the best solution of the respective types. In thin lines the
   solutions bracketing the two sets. The error bars centered at $z=2.5$
   and $z=8.8$ along $Q\Si=1.0$ give the estimated limits for the
   redshifts of complete helium and hydrogen ionizations, respectively
   (see Sec.~\ref{intro}); the vertical dotted black line marks the
   redshift $z=7$ where $Q\nHII$ is found to decrease with increasing
   $z$.\\(In the online journal, the solutions for single and double
   reionization in all the following figures in are in red and green,
   respectively.)}
\label{reion}
\end{figure*}
\begin{table*}
\caption{Values of the adjusted free parameters (besides $\rho\dis=10^6$ \modot\ Kpc$^{-3}$) for the fiducial values of the degenerate
  ones.}\label{bestparam}
\begin{center}
\begin{tabular}{cccccccccccccc}
\hline \hline 
Model & $M\III\lo/$\modot & $\epsilon\B$ & $\epsilon\D$ & $\alpha\G$ & $\epsilon\AGN$ &
$h\rec$ & $f\escG$ & $f\escAGN$\\  
& [0.5,260] & [0,1] & [0,1] & [0,1] & [0,1] & [0,1] & [0,1] & [0,1]\\ \hline
\smallskip
S & $38^{+5}_{-7}$ & $1.00^{+0.00}_{-0.05}$ & $0.00^{+0.05}_{-0.00}$ & $0.33\pm 0.11$ & $0.0016\pm0.0003$ & $0.4\pm 0.3$ & $0.053\pm 0.007$ & $0.0053\pm 0.0006$ \\
\smallskip 
 D & $87^{+9}_{-6}$ & $1.00^{+0.00}_{-0.05}$ & $0.00^{+0.05}_{-0.00}$ & $0.19\pm 0.04$ & $0.0025\pm 0.0004$ & $0.4\pm 0.3$ & $0.055\pm 0.008$ & $0.0053\pm 0.0005$ \\
\hline
\end{tabular}
\end{center}
\end{table*}

\begin{table}
\caption{Fiducial values of the degenerate free
  parameters.}\label{fiducial}
\begin{center}
\begin{tabular}{ccccc}
\hline \hline Model & $\log (Z\crit/$Z$_\odot)$ & $\alpha\III$ & $M\III\up/$\modot\\ & $[-5,-3]$ & [2.15,2.35] & [260,1000] \\ \hline
\smallskip
S  & $-4.6\pm 0.1$ & $2.35^{+0.00}_{-0.10}$ & $300^{+0}_{-40}$\\
D & $-4.0\pm 0.1$ & $2.35^{+0.00}_{-0.10}$ & $300^{+0}_{-40}$\\
\hline
\end{tabular}
\end{center}
\end{table}

To account for all these restrictions parameters $f_2$ and $f_3$ in
the fitting procedure described above must be replaced by the
equivalent ones (for a fixed $\alpha\III$ value) $M\III\lo$ and
$M\III\up$ parameters. Specifically, for some fiducial values of
$Z\crit$ in the range $[10^{-5}$ Z$_\odot,10^{-3}$ Z$_\odot]$, of
$\alpha\III$ in the range [2.25,2.35] (or [2.15,2.35]), and of
$M\III\up$ in the range [260 \modot,500 \modot] (or [260 \modot,1000
  \modot]) warranting that reionization is not aborted, we scan all
possible values of $M\III\lo$ in the range [0.5 \modot,130 \modot],
and of $\epsilon\B$ and $\epsilon\D$ in the range $[0,1]\times[0,1]$,
and adjust in the sequential manner explained above the parameters
$\alpha\G$, $\epsilon\AGN$, $h\rec$, and the two $f\esc$ (see
Fig.~\ref{Flowchart_right}), with the priors on $z\ion$ and $\der_z
Q\nHII$. Using the relations (\ref{deg1}) and (\ref{deg2}), we find
all sets of $Z\crit$, $\alpha\III$, $M\III\up$, and $M\lo\III$ values
leading to every acceptable solution. Lastly, we check whether the
constraints on $\tau\es$ and $y$ are also fulfilled, and whether the
observed galaxy stellar and MBH MFs are well-recovered.

The result will depend, of course, on the somewhat uncertain masses
(130 \modot\ and 260 \modot) delimiting the ranges where \pIII stars
have distinct behaviors. But this is a small drawback compared to the
gain of better constraining $Z\crit$ and the \pIII star IMF.

\section{Results}\label{res}

Following this fitting procedure, we have looked for any possible
solution, in the sense of being compatible with the currently
available data on the high-$z$ Universe and the usual reionization
constraints, and checked whether it has single or multiple
reionization episodes.

The idea of double reionization was advocated by several authors
\citep{Cen03,CFO03,HH03,HHo03,SYAHS04,NC04,FL05} to reconcile the
apparent inconsistency between the value of the redshift of complete
hydrogen ionization obtained from the spectra of distant sources
($z\ion\sim 6$; \citealt{Becker01,DCSM01}) with that found from the
CMB optical depth derived from the {\it First-Year WMAP} data ($z\ion
> 15$ as implied by $\tau\es = 0.17$; \citealt{Ko03}). Subsequent
CMB-based measurements decreased this latter estimate until the
present value of $z\ion$ of about 10. In addition, it became clear
that an {\it instantaneous reionization} event at $z\sim10$ was not
necessarily inconsistent with {\it reionization being completed} by
$z\ion\sim 6$. Consequently, the double reionization scenario lost its
interest. However, the possibility that reionization is double still
remains. As a matter of fact, it is quite natural to expect.

Indeed, \pIII star clusters have a typical mass of $\sim 10^3$ \modot,
which is much lower than the mass $\sim 10^6$ \modot\ of the most
abundant dwarf galaxies at high redshifts (see below). However,
massive \pIII stars emit 40-80 times more ionizing photons than normal
O, B stars, so both kinds of sources compete in the IGM
reionization. What is the dominant contribution? \pIII stars are the
first to form, so they necessarily drive the initial ionization
phase. And, since neutral pristine regions progressively disappear,
the final ionization phase will be driven by normal galaxies. The SFR
of \pIII stars is fully determined by cosmology, and their more or
less top-heavy IMF depends on the unknown value of $M\III\lo$ (for the
fiducial values of $\alpha\III$ and $M\III\up$). Whereas the IMF of
Pop I \& II stars, set by the physics of protostellar clouds, is
well-determined, and their more or less intense SFR depends on the
unknown value of $\alpha\G$. Consequently, the answer to the question
above depends on the relative values of the two independent parameters
$M\III\lo$ and $\alpha\G$. For the $\alpha\G$ value giving good fits
to the SFH of galaxies, the larger the value of $M\III\lo$, the
earlier \pIII stars will begin to ionize the IGM, and the more
probable the double reionization scenario will be.

\subsection{Reionization History}

Figure \ref{reion} shows the reionization histories of the only
acceptable solutions we find. They are concentrated in two narrow
sets, one with one single hydrogen ionization episode at $z\sim 6$,
and the other one with two ionization episodes at $z\sim 10$ and
$z\sim 6$, separated by a short recombination period. The best
solutions of the two kinds, i.e., those that provide the best fit to
the data on the SFH, MBHH, HGMH, IEHs, GAMH, and GASH with the priors
$z\ion=6.0$ and $\der_z Q\nHII(z=7)<0$, are hereafter referred to as
"S" (for single reionization) and "D" (for double reionization).

In all the solutions, the reionization of \HeII also shows two
distinct phases separated by a short recombination period. The first
phase leads to a maximum in the volume fraction of \HeIIp-ionized
regions, although without complete ionization, at $z\sim 10$
in the solutions with double hydrogen reionization, and at $z\sim 6-7$
in the solutions with single hydrogen reionization. The first phase is
also driven by \pIII stars, but the second one, completed at
$z\ionHe\sim 2.8$, is driven by AGN instead of normal galaxies. Thus,
the \HeII reionization history confirms the idea that, when two
independent ionizing sources with distinct timing are at play, it is
natural to have a non-monotonous reionization.

\begin{figure}
\centering
{\includegraphics[scale=.46,bb= 18 170 592 718]{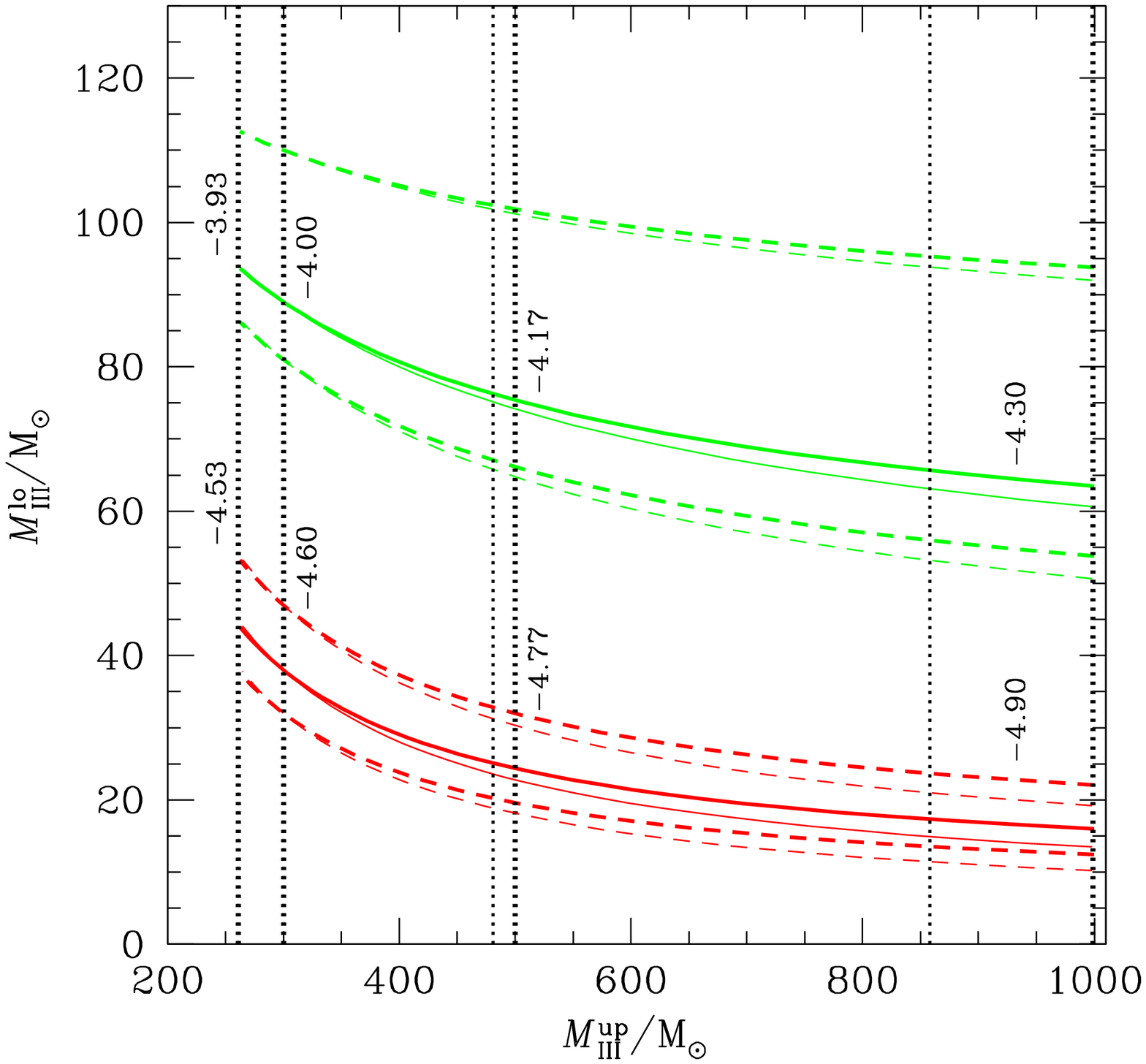}}
 \caption{Degeneracy in $Z\crit$, $\alpha\III$, $M\III\lo$, and
   $M\III\up$ leading to the same solutions with single (lower lines)
   and double (upper lines) reionization (solid lines for the best
   solutions and dashed lines for the respective bracketing solutions)
   for $\alpha\III$ equal to 2.35 (thick lines) and 2.25 (thin
   lines). The vertical black dotted lines mark the maximum possible
   $M\III\up$ values for the quoted logarithms of $Z\crit$ in double
   reionization (upper values), and single reionization (lower
   values).}
\label{degenright}
\end{figure}

\begin{figure}
\centering 
{\includegraphics[scale=.46,bb= 18 170 592 718]{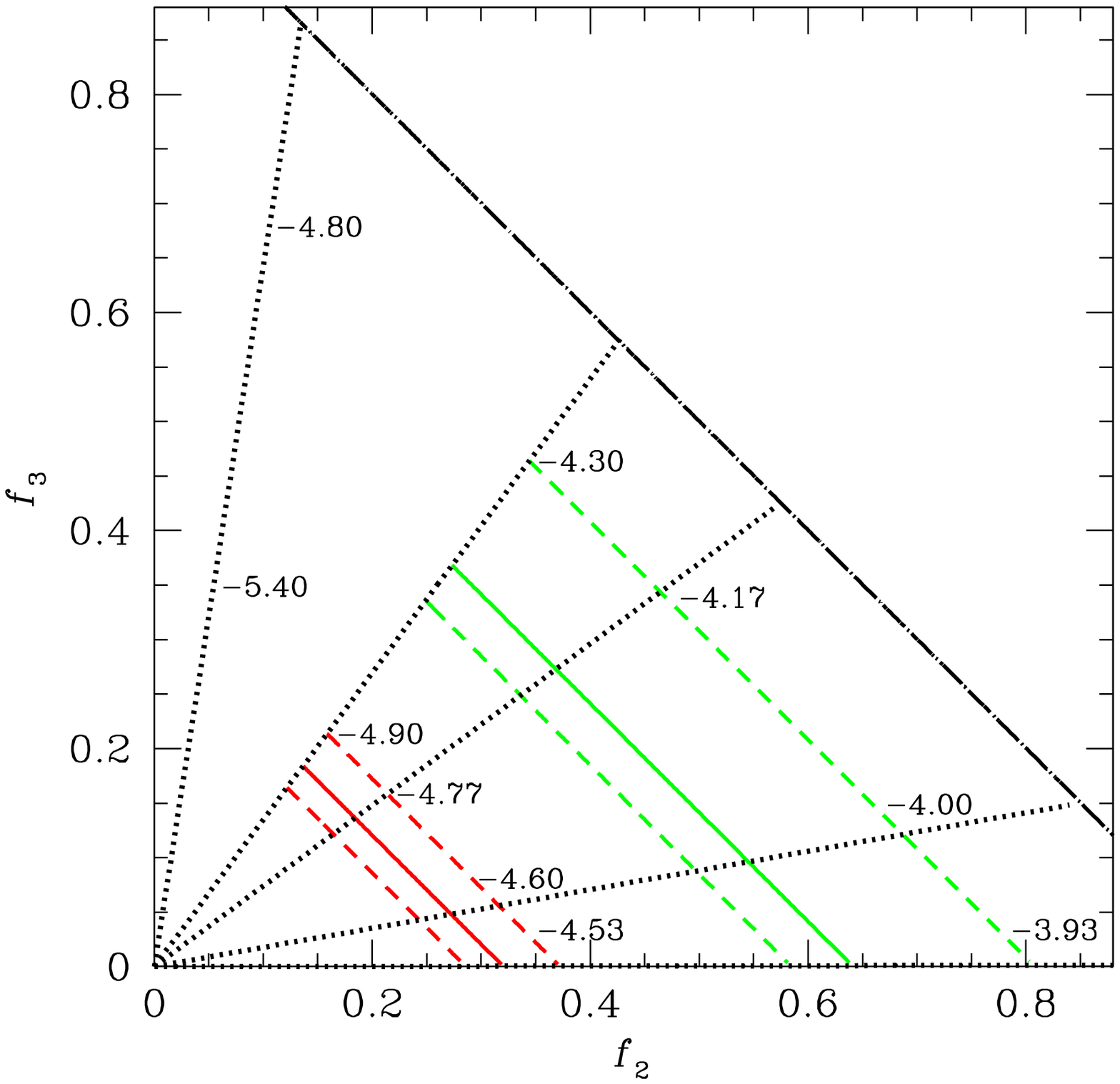}}
 \caption{Same degeneracy (with the solutions for single on the right
   and those for double reionization on the left) as in Figure
   \ref{degenright} in terms of $Z\crit$, $f_2$, and $f_3$. The black
   dot-dashed line, with equation $f_2+f_3=1$, delimits the region
   with possible solutions. The diagonal black dotted lines mark the
   minimum possible $f_2/(f_2+f_3)$ values for the quoted logarithms
   of $Z\crit$ in double reionization (outer values), and single
   reionization (inner values).}
\label{degenleft}
\end{figure}

Table \ref{bestparam} lists the values of $M\III\lo$, $\epsilon\B$,
$\epsilon\D$, $\alpha\G$, $\epsilon\AGN$, $h\rec$, $f\escG$, and
$f\escAGN$ defining models S and D for the fiducial values of
$Z\crit$, $\alpha\III$, and $M\III\up$ given in Table
\ref{fiducial}. The errors in $M\III\lo$ delimit the acceptable
solutions around models S and D; those in $\epsilon\B$ and
$\epsilon\D$ mark the deviations from the best values of these two
parameters still leading to acceptable solutions; and the errors in
the remaining parameters are $1\sigma$ deviations from their best
values in models S and D, estimated from the $\chi^2$-test of their
respective fits. In principle, the degeneracy between $Z\crit$ and the
\pIII star IMF could make the value of $M\lo\III$ to vary from that
quoted in Table \ref{bestparam}. But, for the reasons explained next,
there is actually no much room for the possible values of $Z\crit$,
$\alpha\III$, and $M\III\up$, which should thus be close to those
given in Table \ref{fiducial}.

All the combinations of $Z\crit$, $\alpha\III$, $M\III\up$, and
$M\III\lo$ leading to any given solution can be obtained from one
particular combination leading to that solution and equations
(\ref{deg1}) and (\ref{deg2}) through the relations between the
couples of $M\lo\III$ and $M\up\III$, and $f_2$ and $f_3$ values for a
given $\alpha\III$ value. That degeneracy is shown in Figure
\ref{degenright}. See also Figure \ref{degenleft} for the degeneracy
in $Z\crit$, $f_2$, and $f_3$. Since the minimum possible value of
$f_2/(f_2+f_3)$ for reionization not to be aborted depends on
$Z\crit$, and, for a given $\alpha\III$, $f_2/(f_2+f_3)$ is a function
of $M\III\up$ alone, there is also a one-to-one correspondence between
the maximum possible value of $M\III\up$ and $Z\crit$. Notice that,
for the reasons explained in Section \ref{sources}, the maximum
possible value of $M\III\up$ for each $Z\crit$ is smaller in single
than in double reionization. Specifically, it is equal to the maximum
$M\III\up$ value found in double reionization times 0.85/0.21, where
0.21 and 0.85 are the minimum $f_2/(f_2+f_3)$ values in single and
double reionization, respectively.

As shown in Figure \ref{degenright}, the upper limit of $Z\crit$ set
by the lower limit of $M\up\III$ equal to 260 \modot\ is $10^{-3.93}$
Z$_\odot$ in double reionization, and $10^{-4.53}$ Z$_\odot$ in single
reionization. Whereas the lower limit set by the upper limit of
$M\III\up$ equal to 500 \modot\ (1000 \modot) is $10^{-4.17}$
Z$_\odot$ ($10^{-4.30}$ Z$_\odot$) in double reionization, and
$10^{-4.77}$ Z$_\odot$ ($10^{-4.90}$ Z$_\odot$) in single
reionization. Therefore, the central value and allowed range in double
reionization are: $\log (Z\crit/$Z$_\odot)=-4.0\pm 0.1$ Z$_\odot$,
whereas in single reionization they are: $\log
(Z\crit/$Z$_\odot)=-4.6\pm 0.1$. These are the fiducial values of
$Z\crit$ adopted for the fit. Note that, in the case of double
reionization, the observed properties of the high-$z$ Universe imply a
value of $Z\crit$ of $10^{-4}$ Z$_\odot$, which coincides with that
preferred on pure theoretical grounds \citep{SS05,STSON09,SO10}. Instead,
in the case of single reionization, they imply a value of $Z\crit$
closer to the lower limit of $10^{-5}$ Z$_\odot$. 

In Figure \ref{degenright} we also see that the degeneracy track
associated with any given solution depends slightly on $\alpha\III$,
and so does the exact upper limit of $Z\crit$. But, for all reasonable
values of $\alpha\III$, this dependence is only significant for
unrealistic values of $M\III\up$ close to 1000 \modot. We can thus
adopt $\alpha\III=2.35$ without any loss of generality, even though
the true value of this parameter could be somewhat smaller.

Of course, all $M\III\up$ values smaller than the maximum possible
value for a given $Z\crit$ (i.e. on the left of the corresponding
vertical line in Fig.~\ref{degenright}) together with the $M\III\lo$
value associated with it along the degeneracy track yield essentially
the same solution. (In Fig.~\ref{degenleft} all the couples $f_2$ and
$f_3$ along the degeneracy track upwards that value of $Z\crit$ also
lead to the same solution.) But the maximum possible value of
$M\III\up$ in both single and double reionization, 300 \modot, is
already close to the lower limit of 260 \modot, so the true value of
$M\III\up$ must be close, indeed, to the fiducial one of 300
\modot.

Having fixed the values $Z\crit$, $\alpha\III$, and $M\III\up$, the
central and limiting values of $M\lo\III$ given in Table
\ref{bestparam} have been obtained as follows. In the single
reionization case, the best solutions they define have the central and
limiting values of $z\ion$, equal to $6.0^{+0.3}_{-0.5}$, the
condition $\der_z Q\nHII<0$ at $z=7$ being automatically fulfilled in
this case. In the double reionization case, the best solutions they
define have $z\ion=6.0^{+0.3}_{-0.5}$, but, in addition, the minimum
of $Q\nHII(z)$ is at a redshift larger than, although as close as
possible to $z=7$. This warrants $\der_z Q\nHII<0$ at $z=7$ as wanted,
together with the minimum possible redshift of first complete
ionization, which is very convenient to have the smallest possible
value of $\tau\es$ (see below).

The values of $\epsilon\B$ and $\epsilon\D$ are well constrained by
the observed GAMH and GASH, as expected. In fact, those two properties
appear to be extremely sensitive to small changes in the values of
these (as well as other) parameters. Their values close to one and
zero, respectively, in the two models agree with the results of
analytic and hydrodynamic studies on the effects of SN heating
(e.g. \citealt{DS86,MF99,SS00}), and they are also consistent with
some properties of the nearby Universe such as the high abundance of
$\alpha$-elements in the hot gas of halos or the correlations between
the metallicity of the hot gas and the abundances of early and late
type galaxies in galaxy clusters indicative that most gas outflows
take place from spheroids (e.g. \citealt{S97}).

The values of $\alpha\G$ and $\epsilon\AGN$ are also well constrained
by the observed SFH and MBHH. The larger value of $\alpha\G$ in model
S compensates the lower volume filling factor of ionized regions at
high redshifts in this model compared to model D. Indeed, the SFH is
averaged over the whole space, while stars lie in ionized regions
only. This then leads to a smaller value of $\epsilon\AGN$ in model S
so as to recover the right MBH-to-spheroid mass ratio despite the
larger amount of stars forming in individual spheroids.

\begin{figure}
\centering
{\includegraphics[scale=.44]{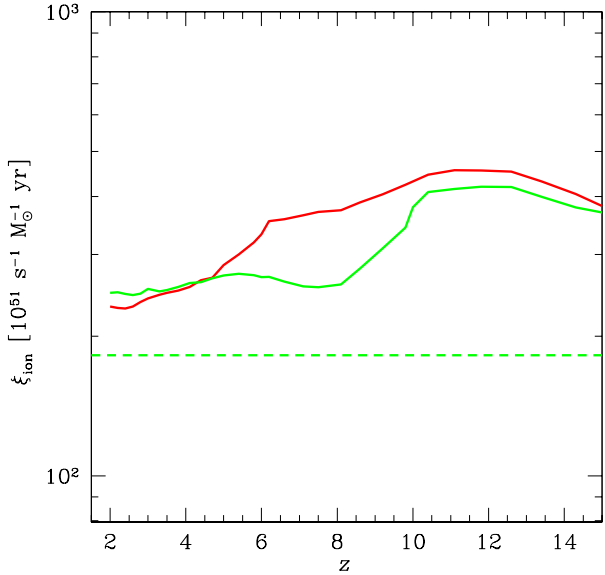}}
\caption{Effective \lyc\ photon production rates predicted in models S
  and D for galaxies with any mass and metallicity (solid lines) and
  for galaxies with stellar masses larger than $\sim 5\times 10^{9}$
  \modot\ and near solar metallicities only (dashed lines; the
  predictions from the two models superpose). The dashed line
  coincides, except for a factor 1.61 due to the different IMF used,
  with the \lyc\ photon production rate predicted in the reionization
  models by \citet{Rea15}, \citet{I15}, and \citet{BouwEtal15}.\\(A
  color version of this figure is available in the online journal.)}
\label{xiion}
\end{figure}

The value of $h\rec$ is the least constrained. The reason for this is
that the data on HGMH show a large scatter, and, even though the data
on the CGMH and STMH are better determined, these cosmic histories are
quite insensitive to $h\rec$. But this result, far from being negative
for our purposes, is heartwarming as the mixing of metals in the hot
gas is hard to model, and the hot gas metallicity estimates are not
very reliable (see Sec.~\ref{observables}).

The values of $f\escG$ and $f\escAGN$ are also well constrained by the
observed IEHs from galaxies and AGN and, more importantly, by the
priors $z\ion=6.0^{+0.3}_{-0.5}$ and $z\ionHe\sim 2.8$. Notice that
the uncertainty of about $\pm 0.8$ in $z\ionHe$
\citep{RGS00,S00,DF09,McQ09,BBHS11} has not been taken into account
since any change of that order in $z\ionHe$ has no significant effect
in the remaining cosmic properties. The values of $f\escAGN$ are one
order of magnitude smaller than those of $f\escG$, which is consistent
with the extra absorption of ionizing photons produced in the
circumnuclear region of galaxies. Regarding the values of $f\escG$,
they agree with the observational estimate (spanning from 0.03 to
0.07) at redshifts out to $z\sim 5$ (see e.g. \citealt{WytEtal10} and
references therein). But these empirical estimates are quite
uncertain, while all recent theoretical studies of reionization based
on the observed SFH find instead $f\escG\ge 0.20$
\citep{Rea15,I15,BouwEtal15}. About half this difference arises from
the fact that those studies neglect the effect of \pIII stars (AGN
have a negligible contribution to the \HIp-ionizing emissivity at
$z>6$), and the other half from the notably larger effective value of
the \lyc\ photon production rate from galaxies, $\xi_{\rm ion}$,
defined through the relation $\dot N\nHII=f\escG\,\xi_{\rm
  ion}\,\dot\rho_\ast$, found in our {\it self-consistent} modeling
compared to that found in those studies (see Fig.~\ref{xiion}). As we
will show in Section 5.3, that increased $\xi_{\rm ion}$ value is due
to a notably larger (by a factor $\sim 2$) contribution of low-mass
galaxies with much lower metallicities (more than one order of
magnitude lower than near-solar).

\begin{figure*}
\centering
 {\includegraphics[scale=1.,bb=198 138 365 726]{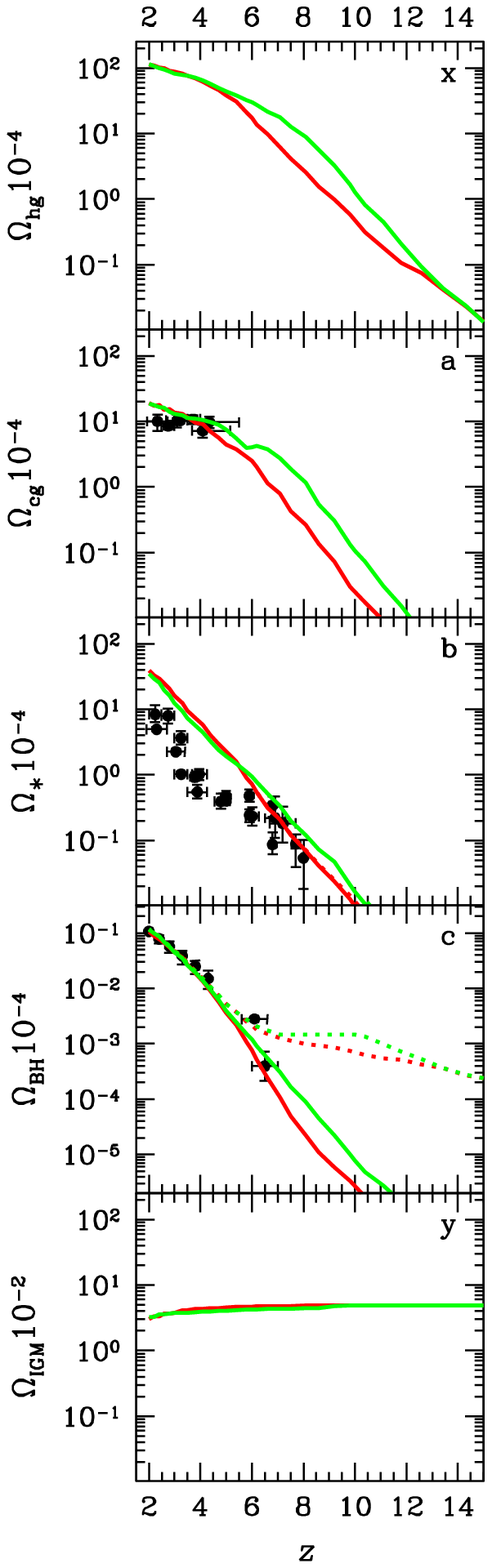}}
 {\includegraphics[scale=1.,bb=196 138 365 726]{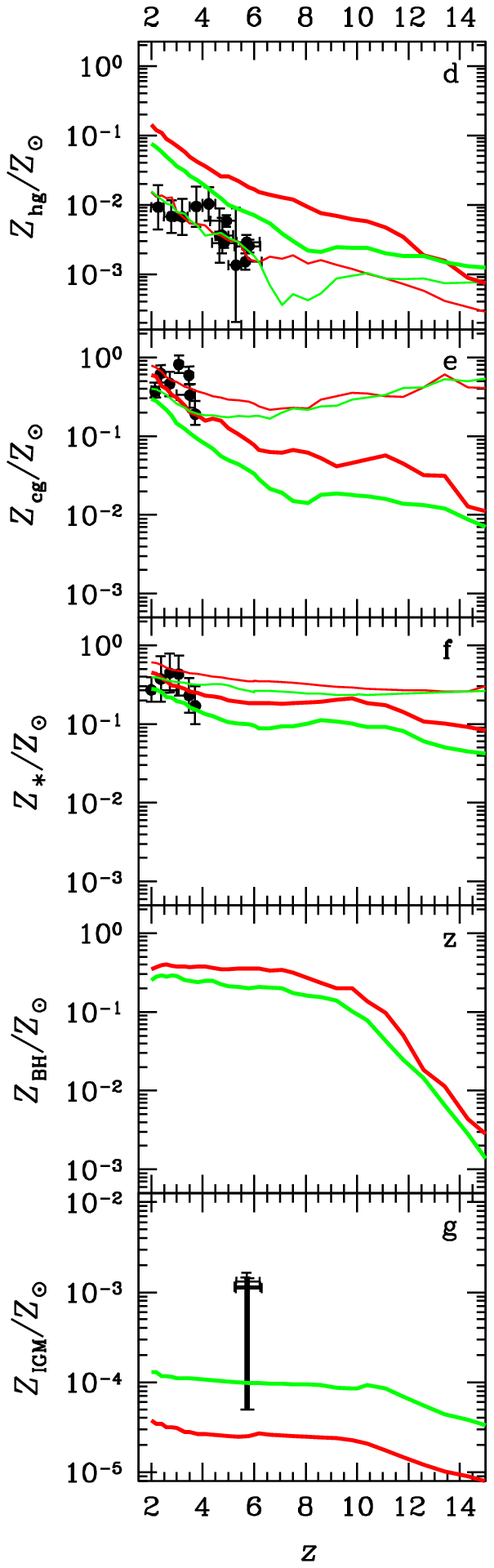}}
 {\includegraphics[scale=1.,bb=196 138 365 726]{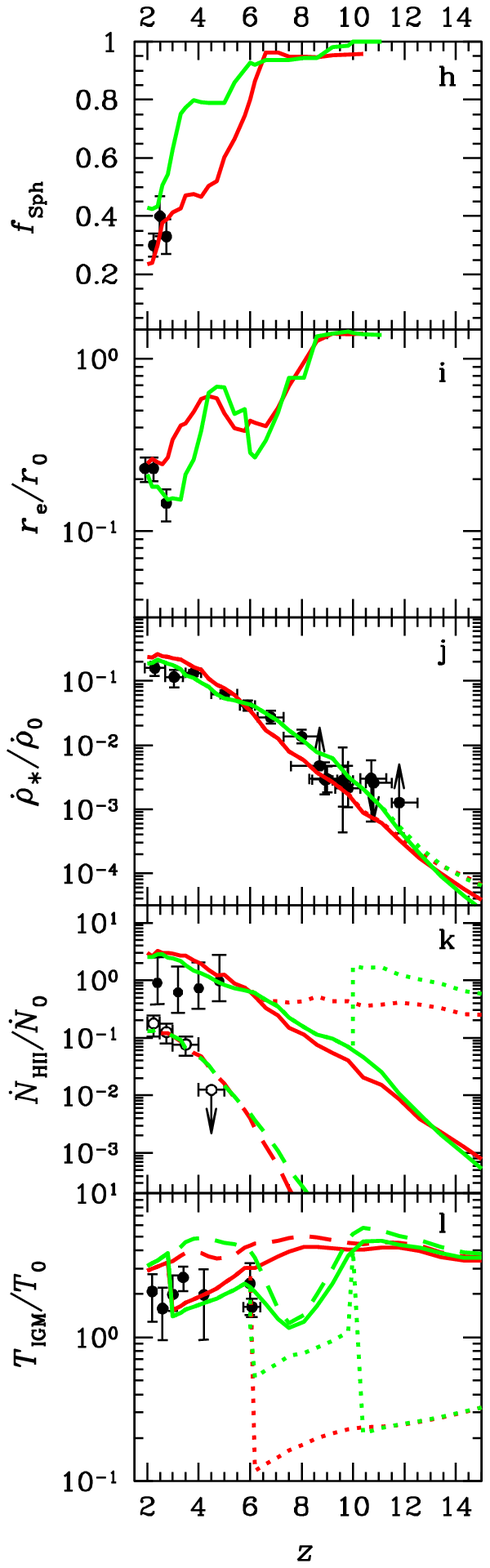}}
\vspace{-10pt}
 \caption{Evolution of galaxies and IGM predicted in models S and D,
   compared to the observational data shown in Figure \ref{data}. We
   plot the contribution from normal galaxies (solid lines) and normal
   galaxies plus \pIII star clusters (dotted lines); in panel (k) we
   also plot the contribution of AGN (dashed lines). In panels (d),
   (e), and (f), we plot the average metallicities in halos and
   galaxies within the mass ranges covered by the data (thin solid
   lines), as well as the average metallicities over all halos and
   galaxies (thick solid lines) with no corresponding data. In panels
   (h) and (i) predictions are only shown at the redshifts where
   galaxies with masses around $10^{11}$ \modot\ are abundant
   enough. In panel (l) we show the average IGM temperatures in singly
   ionized (solid lines), doubly ionized (dashed lines), and neutral
   (dotted lines) regions, the latter shifted a factor 500 upwards at
   $z$ higher than the redshift of first (or unique) complete
   ionization. Note that the sum at any $z$ of the densities in the
   left-hand panels gives the total baryon mass density at that $z$
   (scaled to the constant critical density at $z=0$), and the sum at
   any $z$ of the product of the metallicities in the middle panels
   times the corresponding densities in the left panels gives the
   cumulative mass density of metals produced until that $z$.\\(A
   color version of this figure is available in the online journal.)}
   \label{double}
\end{figure*}

\subsection{Fits to the Observed Global Properties}\label{global}

In Figure \ref{double} we show the fits to the global data that result
in the best models S and D. As it can be seen, both solutions recover
satisfactorily the observed evolution of all cosmic properties. The
only exception is the STH (panel (b)) and, to a much lesser extent,
the CGH (panel (a)). Since the STH is the time-integral of the SFH,
and this latter curve is very well fitted (see panel (j)), the poor
fit to the STH simply reflects the well-known inconsistency between
the datasets of the two cosmic histories. We will comeback to this
point in Section 5.3. The possible origin of the slight excess in the
predicted cold gas mass density at $z<3$ will also be discussed in
Section 5.4.

The fits to the data are particularly good in the case of model D. The
predicted SFH almost fully matches the data, while in model S it only
does marginally (see panel (j)). This is reflected in the respective
$\chi^2$ values of 1.07 and 2.65.\footnote{All $\chi^2$ values
  correspond to the logarithmic curves as plotted in Figure
  \ref{double} and consider only points with well-measured errors bars
  (upper and lower bounds are excluded).} Likewise, the predicted CGH
in model D is almost as shallow as the observed curve in the redshift
range covered by the data, while that in model S is slightly steeper
(see panel (b)); the respective $\chi^2$ values are 1.90 and 2.35. The
predicted HGMH, CGMH, and STMH curves for the metallicities averaged
over all halos and galaxies are also substantially different in the
two models. However, when the averages are restricted to the mass
ranges included in the observations, the two models agree with the
data (see panels (e)--(f)). Similarly, the predicted GAMH and GASH are
somewhat different at high-$z$ in the two models, but they both match
the data at $z\sim 2-3$ (see panels (h) and (i)). Lastly, the
predicted IGMMH is larger in model D than in model S (see panel
(g)). However, the empirical bounds on this quantity are little
reliable (see below). A full comprehensive picture of the
interconnected behavior of all these cosmic histories will be given in
Sections 5.3 and 5.4.

And what about the observational constraints on reionization? In Table
\ref{reionconstraints} we give the values predicted in the best models
S and D. The error bars mark those predicted in the models bracketing
the two sets of acceptable solutions around them. The values of
$z\ion$ and $\der_z Q\nHII$ satisfy, of course, the observed
constraints because they have been enforced as priors. But the values
of $\tau\es$ and $y$ are checked for the first time. The values of the
Compton distortion $y$-parameter predicted in models S and D pass the
test without trouble: they are one order of magnitude lower than the
observed upper limit. The predicted values of $\tau\es$ are also
consistent with the empirical estimates drawn from the WMAP9 and
Planck3 data at the $1\sigma$ level. However, while the value
predicted in model S still agrees at the 1$\sigma$ level with the
estimate of $\tau\es=0.058\pm 0.012$ recently reported by the Planck
team, that in model D only does at the 3$\sigma$ level. We remind that
the way the condition $\der_z Q\nHII(z=7)< 0$ was enforced in this
model warrants that the redshift of first complete reionization is as
small as possible, so the only way to further decrease the predicted
value of $\tau\es$ is to relax that condition. This possibility cannot
be discarded, indeed: the typical outflow velocity of gas clouds in
LAEs could decrease with increasing $z$ so that the absence of LAEs at
$z>7$ would not imply that $Q\nHII(z)$ is decreasing at
$z=7$.\footnote{It could be increasing, and the outflow velocity be
  decreasing rapidly enough to balance that increase.} In any event,
there is also the caveat that the value of $\tau\es$ inferred from the
CMB anisotropies under the instantaneous reionization approximation is
quite uncertain.

\begin{table}
\caption{Reionization constraints.}\label{reionconstraints}
\begin{center}
\begin{tabular}{cccccc}
\hline \hline Model & $z\ion$ & $\der_z Q\nHII(z=7)$
& $\tau\es$ & $y\times 10^{5}$\\ & $6.0^{+0.3}_{-0.5}$ & $<0$ & $0.084\pm
0.017^*$ & $<5$ \\ \hline
\smallskip
S  & $6.0^{+0.3}_{-0.5}$ & $- 0.19^{+0.02}_{-0.02}$ & $0.072^{+0.004}_{-0.005}$ &  $0.25\pm 0.01$\\
D & $6.0^{+0.3}_{-0.5}$ & $- 0.13^{+0.03}_{-0.02}$ & $0.102^{+0.001}_{-0.001}$ & $0.26\pm 0.01$ \\
\hline
\end{tabular}
$^*$This empirical value of $\tau\es$ combines those inferred from the
    {\it WMAP} 9-years and the {\it Planck} three-years data.
\label{ionconst}
\end{center}
\end{table}

\subsection{Checking the Differential Properties}\label{sources}

But the fact that models S and D fit all the observed global
properties, in particular the SFH and MBHH, does not necessarily mean
that they also recover the observed galaxy stellar or MBH MFs. In the
present Section we check that, describing in detail the formation and
properties of galaxies and MBHs of different masses.

At high-$z$ halos are little massive and their concentration is high,
so the hot gas cools very efficiently. But, for this to happen, halos
must first trap gas, that is their virial temperature must be higher than
the IGM temperature, $T\IGM$. The minimum mass of halos trapping gas
at $z$ is 
\beq
M\trap(z)\approx \left\{\frac{[3k\,T\IGM(z)/G]^3}{4\pi\,\Delta(z)\bar\rho(z)/3}\right\}^{1/2},
\label{mtrap}
\eeq
where we have taken into account that the inner mean density of halos
at $z$ is a factor $\Delta(z)$ (close to $178$ at high-$z$) times the
mean cosmic density $\bar\rho(z)$. Therefore, {\it new} luminous
objects (\pIII star clusters or galaxies) will form as long as {\it
  new} halos will trap gas (with undercritical or overcritical
metallicity, respectively), that is as long as $M\trap(z)$ will
decrease or increase less rapidly than the always increasing typical
halo mass of collapse at $z$, $M\crit(z)$, equal to the mass for which
the rms density perturbation equals the critical overdensity for
ellipsoidal collapse at $z$ (e.g. \citealt{JuEtal14b}). $M\crit(z)$ is
fully determined by cosmology, so the formation of new luminous
objects depends only on the evolution of $T\IGM(z)$.

The temperature of the neutral cosmic gas decouples from the CMB
temperature at $z\sim 150$, when Compton heating of residual electrons
from CMB photons becomes negligible. Then, the cosmic neutral gas
cools adiabatically as $(1+z)^2$ until the first \pIII stars form. The
X-ray background they produce (mostly through PISNe explosions) heats
the neutral regions above the CMB temperature by Compton scattering
from the diffuse residual electrons. However, Compton cooling from CMB
photons prevents such a heating from being too marked, and $T\IGM(z)$
diminishes essentially as $1+z$ until Compton cooling progressively
declines by the dilution of CMB photons. Then, the temperature
of neutral regions keeps on diminishing almost as $(1+z)^2$ as X-ray
heating is insufficient to balance adiabatic cooling (see
Fig.~\ref{double}, panel (l)).

Such an evolution of $T\IGM(z)$ in neutral pristine regions implies
that $M\trap(z)$ decreases roughly as $(1+z)^{3/2}$ all the time
except for a short intermediate period where it is kept essentially
constant. Meanwhile, $M\crit(z)$ starts growing very rapidly, and
progressively slows down from $(1+z)^{-2}$ above $z=20$ to
$(1+z)^{-1}$ below $z=6$, being proportional to $(1+z)^{-3/2}$ at an
intermediate redshift of about 10. As a consequence, \pIII star
clusters are permanently forming in neutral {\it pristine}
regions. The situation is different, however, in the neutral {\it
  polluted} nodes that develop in model D during recombination. There,
the temperature is kept much more constant (see Fig~\ref{double},
panel (l)),\footnote{This feature should allow one to tell between
  single and double reionization by looking at the IGMTH. Although
  there is no temperature estimates at $z > 6$, the spin temperature
  of hydrogen atoms is already coupled the neutral gas temperature
  through the Wouthuysen-Field effect, so such a difference between
  models S and D should leave an unequivocal signal in 21 cm line
  observations.} causing $M\trap(z)$ to increase roughly as
$(1+z)^{-3/2}$, that is more rapidly than $M\crit(z)$ since $z\sim
10$. Consequently, no \pIII stars will form in this model after that
redshift.

More specifically, $M\crit(z)$ starts from a very tiny value, and does
not reach what is believed to be the neutralino free streaming mass,
$10^{-6}$ \modot, until $z\sim 23$. This means that the only halos
present at any higher redshift (necessarily more massive than the free
streaming mass of DM particles) are much more massive than the typical
collapse mass at that $z$. Thus their abundance is insignificant. In
fact, the first mini-halo (of $\sim 5\times 10^5$ \modot) able to trap
gas and form a \pIII star cluster (of $\sim 2000$ \modot) within the
{\it current} horizon is found at $z\sim 48$.

As the temperature of the neutral IGM decreases, the mass of
mini-halos able to form \pIII star clusters diminishes, and, at $z\sim
45$, it is low enough for such halos to lose their metal-enriched gas
into ionized bubbles when the most massive \pIII stars explode in
PISNe. Then the IGM metallicity in ionized regions rapidly increases
above $Z\crit$. From that moment, when the ionized metal-enriched gas
is trapped by halos, it rapidly undergoes atomic cooling, and gives
rise to the formation of normal galaxies. But the IGM in ionized
bubbles is also photo-heated to a much higher temperature than in
neutral regions, so halos able to trap ionized gas are again very
massive and rare. In fact, at those redshifts most galaxies form
instead in slightly less massive halos (of $\ga 10^6-10^7$ \modot)
that retained part of the hot gas when \pIII stars exploded. Since the
amount of gas remaining in such halos can be very small, most of those
galaxies are ``failed'' in the sense that they have very low stellar
masses (from $10^2$ \modot\ to $10^5$ \modot).

As halos able to trap {\it fresh} metal-enriched ionized gas become
increasingly abundant, more and more galaxies form with stellar masses
above $10^5$ \modot. This leads to the development of a bimodal
MF. But the formation of normal galaxies causes ionized regions to be
photo-heated to an approximately constant temperature, so $M\trap(z)$
finally increases at the same rate, $(1+z)^{-3/2}$, as $M\crit(z)$ at
$z\sim 10$. Consequently, the formation of new galaxies with stellar
masses above $\sim 10^6$ \modot\ (the lower mass limit is slowly
increasing with decreasing $z$) is halted roughly at that redshift.

From that moment, the evolution of luminous sources differs in the two
models. In model S, \pIII star clusters keep on forming in neutral
pristine regions, and galaxies keep on growing, although not forming,
in ionized ones. While, in model D, \pIII stars stop forming at
$z\sim 10$ when the first complete reionization takes place. (Remember
that no \pIII stars form in the neutral polluted regions that develop
during the subsequent recombination period.) There is thus a sudden
lack of ionizing photons that could definitely interrupt the
reionization process. Fortunately, recombination yields a big dip in
the IGM temperature of ionized regions (see Fig.~\ref{double}, panel (l)) that
allows halos with masses as small as $10^8$ \modot\ to trap fresh
ionized gas. As explained next, the metallicity of this gas is still
overcritical, so an abundant amount of normal galaxies then form which
partially replace \pIII stars in the production of ionizing photons.
This will allow reionization to recover by $z\sim 7.2$.

Indeed, the IGM metallicity {\it in ionized regions} is initially kept
roughly constant due to the fixed proportion of metals ($\propto f_2$)
and ionizing photons ($\propto f_2+f_3$) provided by \pIII stars. But,
once normal galaxies begin to form, there is a larger contribution of
ionizing photons than of metals (remember that metals ejected from
galaxies mainly stay within halos), so the metallicity of ionized
regions begins to decrease. This does not prevent, of course, the {\it
  total average} IGM metallicity from growing (Fig.~\ref{double},
panel (g)) through the increasing weight of ionized regions. But the
metallicity of ionized regions themselves can become
undercritical. This is actually what happens in model S where the
minimum value of $f_2/(f_2+f_3)$ just needs to warrant the
overcritical metallicity when ionized bubbles form. (The undercritical
metallicity of these regions at $z< 10$ in model S has no effect in
the reionization process because no galaxies can form anyway in this
model since $z\sim 10$.) The situation is different, however, in model
D where the IGM temperature decreases since $z\sim 10$. Provided the
minimum value of $f_2/(f_2+f_3)$ is large enough (the maximum value of
$M\III\up$ low enough) to warrant that the IGM metallicity in ionized
regions is still overcritical at $z\sim 10$, new galaxies will form,
indeed, in this model from that redshift to about $z\sim 7$ when the
temperature of the ionized IGM increases again at a high enough rate
for $M\trap(z)$ to grow more rapidly than $M\crit(z)$.\footnote{In
  fact, since the maximum possible value of $M\up\III$ is close to the
  lower limit of 260 \modot, the average IGM metallicity at $z\sim 10$
  cannot be much larger than $Z\crit$ either (see Fig.~\ref{double},
  panel(g)). Therefore, another characteristic of double reionization
  is that the IGM metallicity at $7\la z\la 10$ is close to $Z\crit$.}
The minimum mass of trapping halos in model D during that period is
the largest ever been, and the metallicity of the initial hot gas is
the lowest, so the cooling rate reaches very low values. As a
consequence, a large fraction of these second generation normal
galaxies have very small stellar masses, similar to those of the above
mentioned failed galaxies.

\begin{figure}
\centering
{\includegraphics[scale=.52,bb=50 150 600 700]{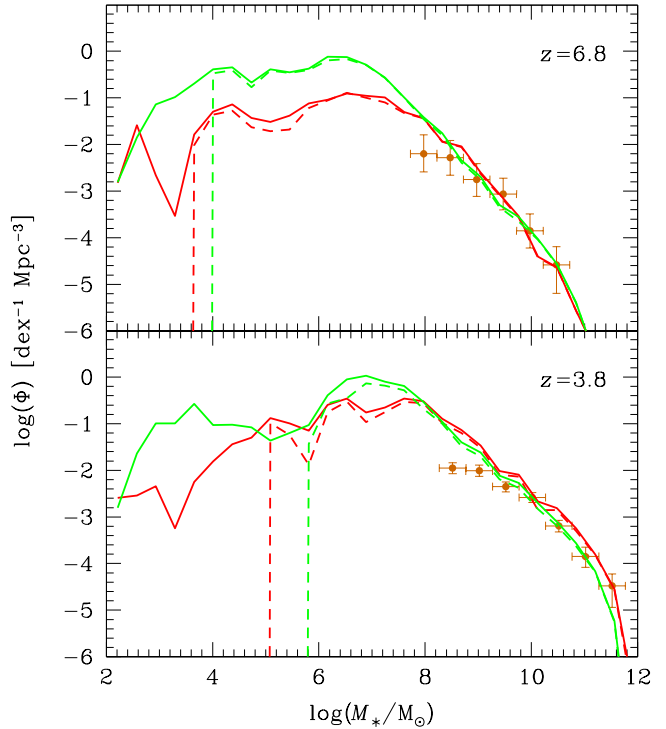}}
 \caption{Galaxy stellar MFs predicted for all galaxies (solid lines)
   and central galaxies (dashed lines) in model S and model D at two
   representative redshifts. Dots with error bars give the MFs
   inferred at those redshifts by \citet{GonzEtal11} from galaxy UV
   LFs, shifted consistently with the stellar mass densities in model
   D (see text).\\(A color version of this figure is available in the
   online journal.)}
\label{Gal_MFs}
\end{figure}

\begin{figure}
\centering
    {\includegraphics[scale=.45,bb= 18 190 592 718]{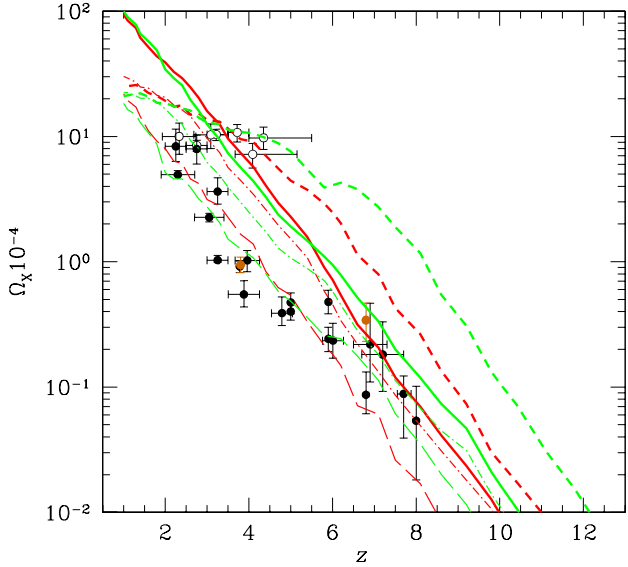}}
 \caption{Same \pIaII STHs (thick solid lines) and CGHs (thick dashed
   lines) as in panels (b) and (a) of Figure \ref{double} but extended
   down to $z=1$. Also shown are the stellar mass densities in disks
   (thin long-dashed lines) and in spheroids (thin dot-dashed lines). To
   avoid confusion the cold gas mass density estimates are here
   plotted as open circles. The stellar mass density estimates
   obtained by \citet{GonzEtal11} from the galaxy stellar MFs shown in
   Figure \ref{Gal_MFs} are plotted as brown dots.\\(A color version
   of this figure is available in the online journal.) }
\label{compBDBH}
\end{figure}

Since $z\sim 10$ in model S or since $z\sim 7$ in model D, no new
galaxies form. Galaxies just grow at the center of halos through the
accretion of new cooled gas and satellite captures. Of course, some of
them freeze out as they become satellites of other central galaxies
when their host halos are captured by other halos. Consequently, at
any $z$ lower than 7, there are in both models galaxies with the whole
spectrum of stellar masses. The only difference is that, while in
model S all the very low mass satellites are failed galaxies, in model
D many of them are second generation galaxies formed at $z<
10$. Notice that, even though halos where those two kinds of galaxies
form have similar masses, $M\crit(z)$ is much greater at $z< 10$ than
at the formation of failed galaxies, so those second generation
low-mass galaxies are more abundant than failed ones.

Figure \ref{Gal_MFs} shows the galaxy stellar MFs that result in
models S and D at redshifts $z=3.8$ and $z=6.8$. To compare them with
the empirical MFs inferred by \citet{GonzEtal11} at the same redshifts
we need to account for the above mentioned inconsistency between the
SFH and STH datasets. Indeed, at high- and low-$z$, the predicted STH
curve, consistent with the empirical SFH one, is close to the upper
envelope of the stellar mass density estimates, and between $z\la 6$
and $z\sim 4$ it is even higher (see
Fig.~\ref{compBDBH}). \citet{WTH08} put forward that such an
inconsistency could be due to a $z$-varying IMF. However, the largest
discrepancy is found at an intermediate redshift range, which would
require the IMF to have a strange non-monotonous trend. A simpler
explanation is that the correction for dust attenuation achieved in
the galaxy UV luminosities is typically underestimated. Galaxies in
the redshift range $4\la z\la 6$ would be substantially brighter than
believed, the effect being less marked at higher and lower redshifts
because of the lower metallicity of galaxies and the more accurate
correction achieved, respectively. But, whatever the real cause of
that inconsistency is, provided all the observed galaxies are
similarly affected, the {\it shape} of the MFs derived from the UV LFs
would be right except for an horizontal shift. Specifically, according
to the stellar mass density estimated inferred by \citet{GonzEtal11}
at redshifts 3.8 and 6.8 and the theoretical STH curve consistent with
the empirical SFH one, the MFs used by \citet{GonzEtal11} to infer
those estimates should be shifted in model D (S) by a factor of about
6.0 (7.9) and 1.3 (0.8), respectively. Taking into account that the
extrapolations of those empirical MFs were underestimated by factors
$1.6$ (1.3) and $2.1$ (1.4), respectively, the actual shifts to be
applied at those redshifts are 9.5 (10.3) and 2.7 (1.2),
respectively. As it can be seen in Figure \ref{compBDBH}, these shifts
make the empirical MFs fully match indeed the predicted ones except
for the incomplete low-mass bins (see below). Thus the predicted
galaxy stellar MFs are in very good agreement with the
observations. The fact that the MF predicted at $z=3.8$ also matches
the Schechter fit to the original (with no shift) galaxy stellar MF
recently derived by \citet{Cap15} at $3\le z\le 4$ assuming a constant
IMF also gives strong support to the interpretation that the
inconsistency between the SFH and STH datasets is due to a deficient
correction for dust attenuation. According to our results in model D
(S), the UV luminosities of galaxies inferred by \citet{GonzEtal11} at
$z=3.8$ and $z=6.8$ would be underestimated by 2.5 (2.7) mag and 1.0
(1.1) mag, respectively.

Regarding the behavior at lower-masses of the predicted MFs, we find
the turnover at $\sim 3.5$ dex from the knee in model S, and at $\sim
4.5$ dex in model D. These results are consistent with the latest UV
LFs of galaxies observed at $z\sim 3$ \citep{Lea16,A16}, which find
that the turnover is not yet detected at $\sim 6.5$ mag (or $\sim 2.6$
dex in mass) from the knee. We remark that all previous models of
galaxy formation predicted the turnover at just $\sim 3.5$ mag ($\sim
1.4$ dex in mass) from the knee \citep{J13,K13}, which is contradicted
by these observations. In this sense, our model is the first to
predict MFs consistent with the observations. According to it, such
LFs should still go $\sim 2.25$ mag or $\sim 4.75$ mag deeper in
models S or D, respectively, for the turnover to be finally detected.

Model D also predicts the existence of a second bump at very low
masses in the galaxy stellar MFs at $z=3.8$ caused by the second
generation low-mass galaxies that become satellites soon after their
formation. This bump is far from the current observational limits of
the galaxy stellar MFs at those redshifts. However, given the small
mass of those satellites, a large fraction of them should still
survive at $z=0$. In model S, there is no such a bump, although the
abundance of very low mass galaxies is also quite high. Thus a
prediction of our model, regardless of the version, is that it should
be possible to detect at $z=0$ very old satellites with masses between
$10^2$ \modot\ and $10^6$ \modot. We describe next the expected
metallicity, morphology, and size of those predicted relics.

The hot gas metallicity increases in parallel to the cold gas and
stellar metallicities (see Fig.~\ref{double}, panel (d), (e) and
(f)). The only effect that partly balances that increase is the
accretion by halos of low-metallicity IGM in ionized regions. That
increase is more marked in model S than in model D due to the steeper
SFH caused by the larger $\alpha\G$ value, and hence, the larger
amount of stars that are permanently forming in galaxies. Another more
subtle difference between the two models is the small dip in the hot
gas metallicity in model D after $z\sim 10$ due to the large amount of
fresh gas that is being incorporated by halos during recombination
owing to the dip in the temperature of the ionized IGM. This
translates, of course, in less marked dips in the CGMH and STMH of
model D. The stellar and cold gas metallicities of the second
generation galaxies formed in this model during recombination, similar
to those of the primordial failed galaxies formed in both models, are
of about 10$Z\crit$, i.e. significantly lower than
average.\footnote{These metallicities are nonetheless larger than the
  metallicity of the initial gas due to the rapid increase in the
  metallicity of the gas produced during star formation.} On the
contrary, the metallicities of galaxies more massive than $5\times
10^9$ \modot\ are, of course, larger than average (see also
Fig.~\ref{double}, panels (e) and (f)) and show a well-understood
behavior with $z$: at the highest redshift where those galactic masses
are assembled, such galaxies are extraordinarily massive, having been
formed in the merger of evolved high metallicity galaxies; then the
metallicities slowly decrease as the masses become increasingly
ordinary and such galaxies form through other channels (including
accretion of hot gas with lower metallicity); and, when the masses of
the galaxies approach the average galaxy mass at $z\sim 4$, they begin
to increase just as the average metallicities do since $z\sim 6$.

The morphology of galaxies follows from the following general
trends. When a galaxy forms, the gas it collects comes from the very
central part of the halo, so it has little angular momentum, and it is
pilled up in a spheroid. Consequently, the failed galaxies in models S
and D as well as the second generation low-mass galaxies in model D
are spheroids. But, soon after their formation, the morphology of
galaxies evolves at the center of halos. At very high-$z$, when
cooling is very efficient, galaxies usually collect large amounts of
cold gas that render disks unstable. In addition, halos are small, and
galaxy mergers frequent, so, whenever disks are stable, they are
short-lived and never reach large radii. But, as halos grow, cooling
becomes less efficient and galaxy mergers less frequent, so disks grow
larger and are more long-lived. As a consequence, the fraction of
spheroid-dominated galaxies slowly decreases with decreasing $z$. On
the other hand, at any given $z$, the most massive galaxies tend to be
spheroids arising from galaxy mergers of massive objects, while disks
preferentially lie in intermediate mass galaxies. These trends explain
the observed fraction of spheroid-dominated objects with masses above
$\sim 10^{11}$ \modot\ at $z\sim 2-3$. We remark, however, that about
$25$\% of galaxies at those redshifts are red dead ellipticals with no
star formation. These galaxies have been excluded from the results
shown in panel (h) of Figure \ref{double}, as they would go
undetected. Therefore, the real fraction of spheroid-dominated massive
galaxies at those redshifts is $\sim 0.65$ rather than $\sim 0.40$.

The size of galaxies increases in parallel to their mass since the
beginning. In fact, when they form as spheroids from a gas cloud with
small angular momentum, even though the metallicity of the gas is low
(about $Z\crit$), they undergo a large dissipative contraction until
they reach the density of molecular clouds. This leads to stellar
systems with the typical density of globular clusters. The size of
more evolved spheroids depends on the amount and metallicity of the
gas they collect (see eq.~[\ref{radius2}]). The result is an evolution
of the average equivalent radius (measured as the square root of the
mass-weighted squared equivalent radii of the bulge and the disk, if
any) of spheroid-dominated galaxies with masses above $10^{11}$
\modot\ that recovers the observations at $z\sim 2-3$
(Fig.~\ref{double}, panel (i)). Specifically, when those very massive
galaxies appear for the first time in dry mergers of massive
spheroids, their equivalent radius is even slightly larger than that
of present day slightly contracted spheroids of the same mass; this
radius begins to decrease as the mass becomes increasingly ordinary
and those galaxies form from unstable disks and wet mergers; it then
increases again up to a secondary maximum caused by the increasing lack of gas
followed by the increase of the gas metallicity at $z\sim 5$ (see
Fig.~\ref{double}, panel (e)); and it finally increases until $z=0$
due to the definite small amount of gas present in galaxies.

Thus the properties of very low mass galaxies are essentially the same
in both models. They are among the first objects with \pIaII stars to
form; they are spheroids with no gas; they are very compact, with the
typical density of globular clusters; and their \pIaII stars have very
low metallicities (of about 10--100 times $Z\crit$, with $Z\crit$
equal to $\sim 0.25 \times 10^{-4}$ Z$_\odot$ in model S, and $\sim
10^{-4}$ Z$_\odot$ in model D). On the other hand, those of them that
froze out as satellites soon after formation have very large
baryon-to-DM mass ratios as their original DM halos are truncated by
the potential well of the also quite dense capturing halo near the
region where all the stars are strongly concentrated. And, the
fraction of those satellites that will be captured long time after
(because of the slight dynamic friction they suffer) by evolved, much
less dense galaxies will survive inside them as small independent
dynamical entities.

Clearly, all these properties greatly resemble those of present day
globular clusters. Since such objects are often seen to freely orbit
within halos, they should have formed outside galaxies and have a
cosmological origin similar to them. It is thus not surprising that a
self-consistent model such as AMIGA that monitors the formation and
evolution of luminous objects since the dark ages does not miss
them. Notice that, if this is indeed the origin of globular clusters,
the larger abundance of such objects compared to massive galaxies
is more consistent with model D than with model S.

\begin{figure}
\centering
    {\includegraphics[scale=.45,bb=18 190 592 718]{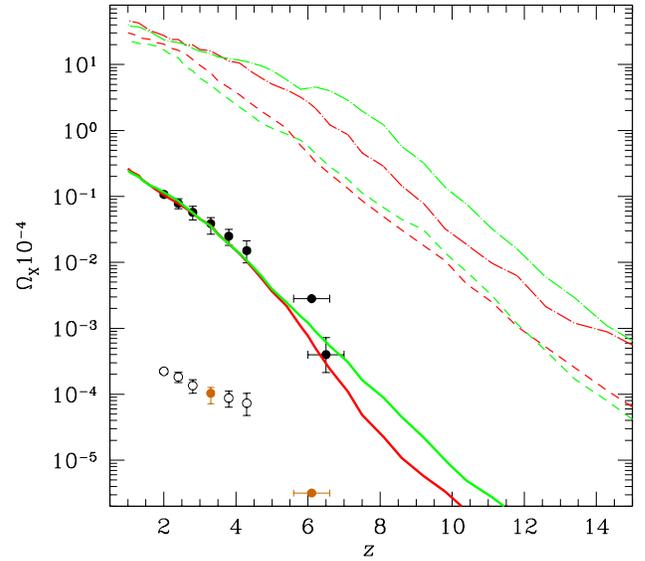}}
 \caption{Same MBHHs (thick solid lines) as in panel (c) of Figure
   \ref{double} but extended down to $z=1$. Also shown are the
   spheroid mass density history (thin dashed lines), and disk mass
   density history (thin dot-long dashed lines). Empty circles give
   the MBH mass density estimates obtained from rough (incomplete and
   uncorrected for obscuration and inactivity) MBH MFs. Brown dots
   at $z=3.3$ and $z=6.1$ correspond to the estimates derived by
   \citet{Kellea10} and \citet{WillEtal10b}, respectively, from the
   MFs used in Figure \ref{MBH_MFs}. \\(A color version of this figure
   is available in the online journal.)}
\label{BHDaS}
\end{figure}

\begin{figure}
\centering
{\includegraphics[scale=.52,bb=50 150 600 700]{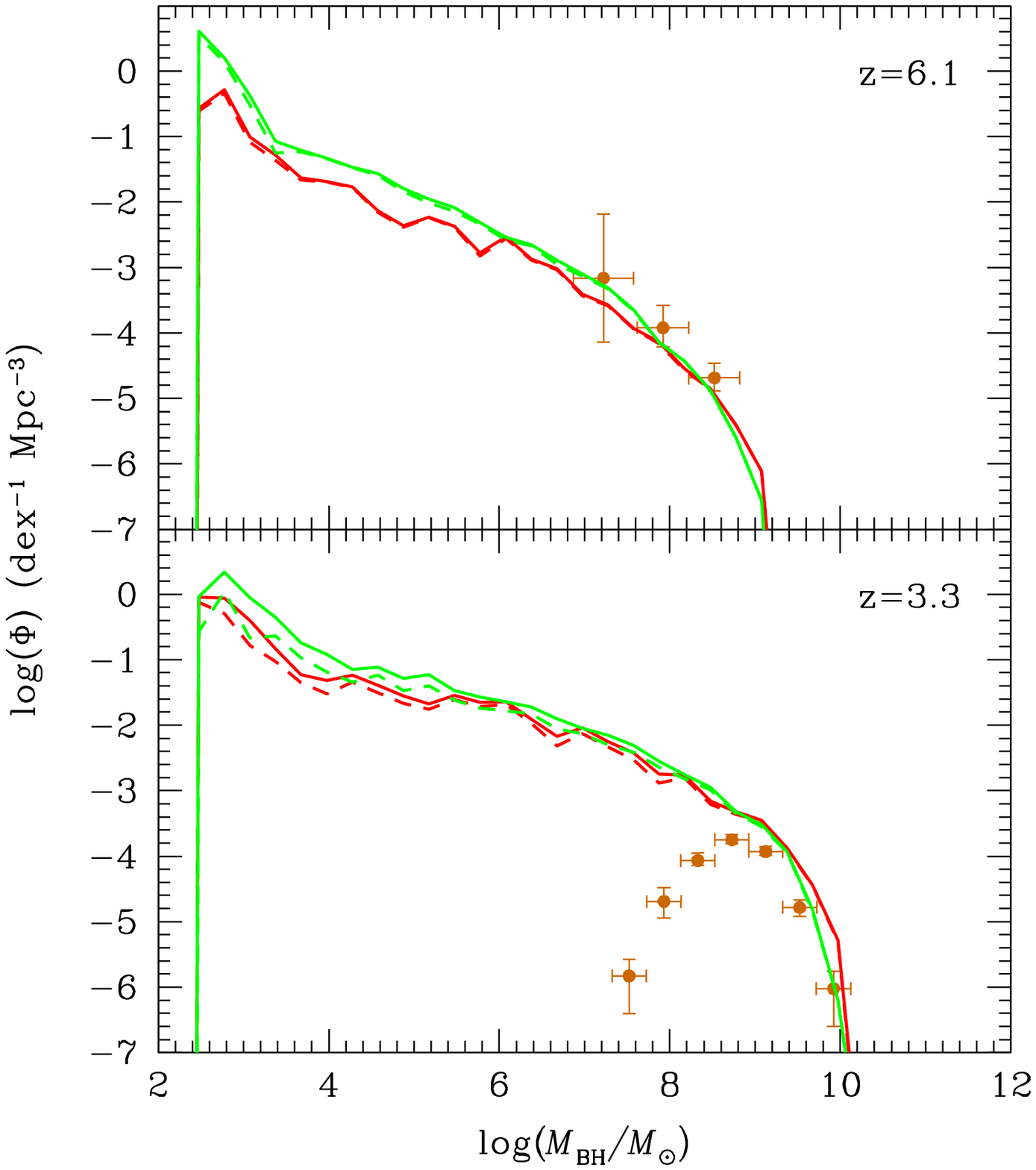}}
 \caption{MFs of MBH in central galaxies (dashed lines) and in all
   galaxies (solid lines) predicted in models S and D at two
   representative redshifts. Dots with error bars give the MBH MFs
   inferred at those redshifts by \citet{WillEtal10b} (top panel) and
   \citet{VeEtal} (bottom panel) from optical AGN LFs, shifted
   consistently with the MBH mass densities in model D and a
   well-suited $f_{\rm FWHM}$ value (see text).\\(A color version of
   this figure is available in the online journal.)}
\label{MBH_MFs}
\end{figure}

Let us now turn to the MFs of MBHs. MBHs grow since the remnant
mini-MBHs of \pIII star clusters begin to develop at the center of
galaxies (Fig.~\ref{double}, panel (c)). There, they merge with the
MBHs brought by accreted or merged galaxies having reached the center
of the galaxy by dynamical friction, and accrete new material. Indeed,
when some amount of gas reaches the spheroid it fuels the MBH. The
associated AGN is then lit, and begins to reheat the circumnuclear
gas, quenching the star formation that is going on in the
spheroid. The more massive the MBH, the larger both the AGN bolometric
luminosity and the fraction of gas reheated and expelled from the
spheroid back into the halo. This leads to a self-regulated growth of
MBHs and stellar spheroids, which tends to a constant MBH-to-spheroid
mass ratio as observed (Fig.~\ref{BHDaS}). This thus gives strong
support to the mechanism proposed by \citet{DiM05} included in AMIGA
for the coupled growth of MBHs and spheroids.

The parallel evolution of spheroids and MBHs goes together with that
of disks. Of course, when gas replenishment begins to decline by
$z\sim 7$ (see Section 5.4), the disk and spheroid mass density
histories become slightly shallower. (The change of slope is a little
less marked in spheroids than in disks due to their unchanged behavior
in major mergers.) But this has a minor effect in the parallel
evolution of the spheroid and MBH stellar mass densities because
mergers become at the same time increasingly dry. Indeed, in dry
mergers spheroids grow through the addition of stars belonging to the
merging galaxies, while MBHs grow by coalescence of the respective
MBHs and, since the MBH-to-spheroid mass ratio is the same in all the
progenitors, it remains unchanged in the final object. As a
consequence, the MBH and spheroid mass ratio will be kept essentially
unaltered until $z=0$.

Figure \ref{MBH_MFs} shows the MBH MFs predicted at redshifts $z=3.3$
and $6.1$. Like in the case of the galaxy stellar MFs, to compare
these MFs to the empirical ones derived at the same redshifts by
\citet{WillEtal10b} and \citet{VeEtal} we must take into account any
effect that could affect the latter. As mentioned, the observed AGN
LFs are greatly affected by incompleteness, obscuration, and
inactivity, which cause the MBH mass density estimates derived from
them to be much lower than expected from the observed constant
MBH-to-spheroid mass ratio. Incompleteness affects in an uneven way
the different mass bins of the MBH MFs, but their large mass end
should be complete. Obscuration is also larger for low-mass objects
\citep{UeEtal03}, but the most massive bins should be similarly
affected. Thus a vertical shift by a factor $s_1$ of a few hundreds
should be enough to correct the empirical MBH MFs for obscuration and
inactivity. In addition, there is the uncertainty in the MBH mass
estimates themselves coming from the poorly determined virial factor,
$f_\sigma$ or $f_{\rm FWHM}$. For instance, the MBH masses derived by
\citet{WillEtal10b} and \citet{VeEtal} at $z=6.1$ and $3.3$,
respectively, from emission-line reverberation assume $f_{\rm
  FWHM}=1.17$ \citep{C06}, while, according to \citet{Y16}, this
factor could be up to a factor 5 smaller. Consequently, the empirical
MBH MFs derived by those authors might also need to be shifted
horizontally towards large masses by a factor $s_2$ in the range $1\ge
s_2\ge 0.2$. The value of $s_2$ is unknown, but it should be
independent of $z$. On the other hand, the shifts $s_1/s_2$ these
effects would cause to the MBH mass density estimates derived by
\citet{Kellea10} and \citet{WillEtal10b} should make them match the
predicted MBHH curve consistent with the observed MBH-to-spheroid mass
ratio, equal to $\sim 400$ and $\sim 250$ at $z=3.3$ and $z=6.1$,
respectively (see Fig.~\ref{BHDaS}).\footnote{These values essentially
  hold for both models D and S.} Taking into account the different
degrees of completion in the extrapolated MFs used in the calculation
of those MBH mass density estimates, the vertical shifts to be applied
to the corresponding MFs at $z=3.3$ and $z=6.1$ should be 1.58 and
1.70 times larger, respectively. Therefore, with the only freedom of
the $z$-independent value of $s_2$ (with $1\ge s_2\ge 0.2$), it should
be possible to shift the empirical MFs at the two redshifts so as to
match those predicted at the two redshifts. As shown in Figure
\ref{MBH_MFs}, this is indeed what we find for $s_2=0.33$ (or $f_{\rm
  FWHM}=0.39$). Notice that, as a consequence of the small correction
in $f_{\rm FWHM}$, the upper mass limit of MBHs at $z=3.3$ becomes
$\sim 10^{10}$ \modot, which is more reasonable than the original
value of $\sim 3 \times 10^{10}$ \modot\ if we compare with such an
upper limit at $z=0$.

\subsection{Reliability of the Model}

It might be argued that the model could be wrong and the solutions
found for some appropriate values of the free parameters still recover
the observed global and differential cosmic properties at $z>2$. But
such good results have been achieved with {\it less than one degree of
  freedom per curve}, which is unbelievable for any model that would
not include the right physics. To see this we must simply think, for
instance, in the fitting by means of a well-suited function with no
physics of {\it one single monotonous curve} such as the observed SFH
at $z>2$ that requires four parameters (e.g. \citealt{Rea15}).
Furthermore, not only does our model with only 9 degrees of freedom
provide very good fits to the dozen observed curves at $z>2$ (plus the
galaxy stellar and MBH MFs), but it also automatically recovers all
their extreme and saddle points at $z\sim 2$ with no need to change
the value of any parameter.

As mentioned, disks become increasingly abundant as $z$ decreases, so
the cold gas mass density increases (Fig.~\ref{double}, panel
(a)). But cooling becomes increasingly inefficient, and by $z\sim 7$
the rate of gas storage in disks begins to decline and the CGH becomes
increasingly shallower. This effect is particularly marked in model D,
where the CGH reaches a plateau between $z\sim 5$ to $z\sim 3$. The
observed CGH is also flat at those redshifts, but the plateau reaches
$z\sim 2$ (see Fig.~\ref{double}, panel (b)). There are several
possible causes for that small discrepancy, such as the fact that the
empirical cold gas mass density estimates do not include molecular gas
or that observations may be biased towards large (massive) DLAs which
have the lowest gas content. But the most likely reason is that
satellite interactions, neglected in the present work, begin to have a
significant impact in the amount of gas present in disks. In any
event, the predicted stagnation of cold gas around $z\sim 4$
corresponds to a saddle point, not to a maximum, in agreement with
observation (e.g. \citealt{PMSLI03,PHFW05}). At lower redshifts the
temperature of the singly and doubly ionized IGM begins to diminish
(see Fig~\ref{double}, panel (l)), causing a rather constant slope in
the HGMH, and the continuation of cooling.

\begin{figure}
\centering
    {\includegraphics[scale=.45,bb=18 190 592 718]{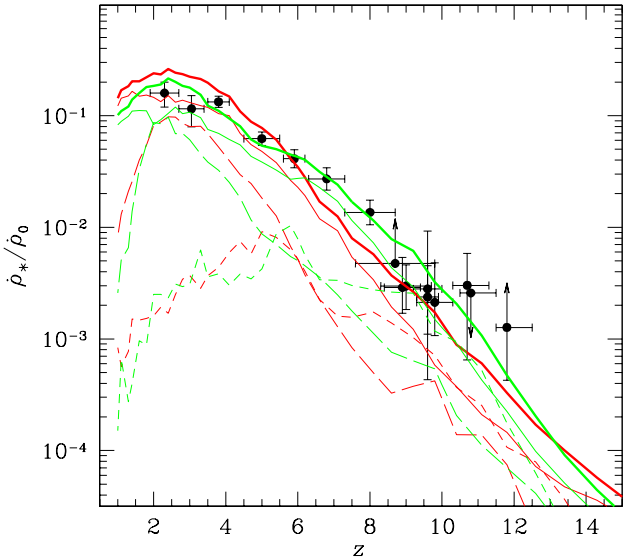}}
 \caption{Same \pIaII SFHs (thick solid lines) as in panel (i) of
   Figure \ref{double} but extended down to $z=1$. Also shown are the
   contributions from disks (thin solid lines) and spheroids arising
   from galaxy mergers (thin short dashed lines) or direct infall of
   cold gas in unstable disks (thin long-dashed dashed lines). \\(A
   color version of this figure is available in the online journal.)}
\label{new}
\end{figure}

Of course, the SFR density of ordinary (\pIaII) stars increases in
parallel to the cold gas mass density (see Fig.~\ref{compBDBH}), so,
when gas replenishment begins to decline by $z \sim 7$, the SFR
density also slows down. Until that redshift, spheroids were much more
abundant than disks, which never collected large amounts of
gas. Nonetheless, disks were small, so their dynamical time was short,
and the SFR density in disks was notable (see Fig.~\ref{new}). The
small amount of gas in disks also prevents the SFR density in mergers
from being too large. It is just comparable to that in unstable disks
at high-$z$, and reaches a maximum at $z\sim 6$. This does not mean
that mergers are less frequent from that moment; they just become
increasingly dry. As the cooling rate declines, the SFR in {\it
  individual} unstable disks also does, but the corresponding {\it
  global} SFR density keeps on increasing because of the increasing
abundance of spheroids contributing to it. The same happens with
disks: even though the size of disks increases and the amount of cold
gas they harbor decreases with decreasing $z$ so that the SFR in
individual disks decreases, the global SFR density in disks keeps on
increasing because of the increasing abundance of disks. As a result,
the ratio between the SFR densities in disks and spheroids is kept
roughly constant. However, the declining amount of cold gas in disks
ultimately causes the SFR in both unstable and stable disks to reach a
maximum at $z\sim 2$, and then to rapidly diminish towards $z=0$ (see
Fig. 12). Therefore, the origin of this maximum in the SFH is the same
as for the plateau in the CGH: the lack of gas replenishment. However,
in the SFH case, such a stagnation leads to a real maximum, not to a
saddle point like in the CGH case. The drop in the SFR in disks and
spheroids at $z\sim 2$ is so marked that, even though at lower redshifts
the cold gas mass density begins to increase again, the total SFR
density will keep on decreasing until $z=0$.

\begin{figure}
\centering
    {\includegraphics[scale=.45,bb=18 190 592 718]{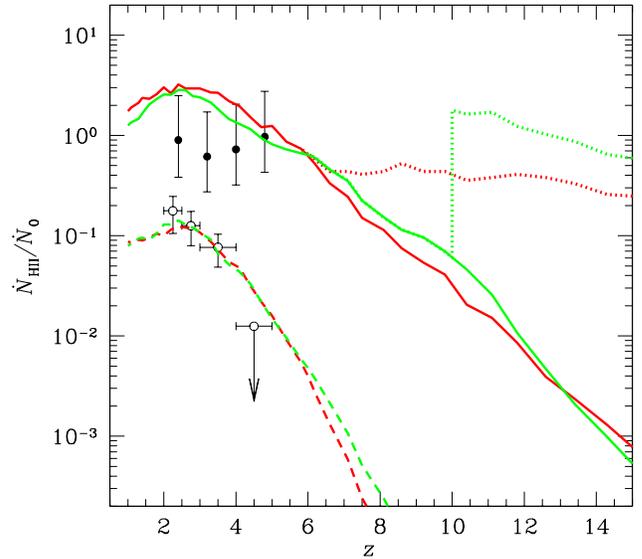}}
 \caption{Same IEHs from normal galaxies (solid lines) and AGN (dashed
   lines) as in panel (k) of Figure \ref{double} but extended down to
   $z=1$. \\(A color version of this figure is available in the online
   journal.)}
\label{renew}
\end{figure}

The stagnation of cold gas mass at $z\la 5$ has a similar effect on
the SFH in models S and D. But the rest of the curve is slightly
different in the two models. In model D, the SFH grows more steeply at
high-$z$ despite the smaller value of $\alpha\G$ (see
Fig.~\ref{double}, panel (j)) because of the more rapid growth of
ionized regions where normal galaxies lie. But, near $z=10$ the
formation of \pIII stars declines and ionized regions stop growing,
causing the SFH to become shallower. This yields a concave curvature
in the SFH of model D that perfectly fits the data. Instead, the SFH
in model S has a more constant slope. This difference in the SFH of
both models combined with the large abundance of spheroids with a very
efficient SN heating also translates into a larger amount of hot gas
in model D between $z\sim 12$ to $z\sim 5$ (see Fig.~\ref{double},
panel (x)). Unfortunately, there are no observational data on this
quantity yet, so the behavior of the HGMH in the two models cannot be
checked.

The behavior of the IEH from galaxies and AGN follows from the
evolution of those two kinds of sources described above. At very
high-$z$, \pIII stars have the dominant contribution. But this
contribution diminishes as neutral pristine regions progressively
disappear. In model D it completely vanishes at $z\sim 10$ when \pIII
stars stop forming, whereas in model S it continues until $z\sim
6$. On the other hand, the maximum contribution from AGN is one order
of magnitude smaller at $z\sim 2$. Consequently, the \HIp-ionizing
emissivity at $z\la 6$ is always dominated by normal galaxies, and its
evolution closely follows the SFH associated with ordinary stars. In
particular, the IEH also reaches a maximum at $z\sim 2$ (see
Fig.~\ref{renew}). What is less obvious is the maximum in both models
S and D of the \HIp- and \HeII-ionizing (as well as the X-ray)
emissivities from AGN at that redshift (Fig.~\ref{renew}): the MBH
mass density is ever increasing at $z<2$ (see Fig.~\ref{BHDaS}) so we
could naively expect their emissivities to also do. What causes these
maxima is the fact that the most massive MBHs stop being fueled by
that redshift (their galaxies undergo dry mergers). Consequently, the
bright AGN stop radiating as quasars, and remain in the more quiescent
radio phase. This has a major impact in the global AGN emissivities
because bright AGN have the largest bolometric luminosity
(e.g. \citealt{Hea03}). However, massive MBHs keep on growing through
dry mergers and less massive ones through wet mergers of low-mass
galaxies that still harbor gas.

\section{Summary and Discussion}\label{conclude}

Observational data on the high-$z$ universe are nowadays numerous and
accurate enough to severely constrain the interconnected evolution of
galaxies and IGM. In fact, they would completely determine it provided
the IGM metallicity and temperature were known at $z\ga 6$. These
pieces of information are needed, indeed, to determine the threshold
metallicity $Z\crit$ for atomic cooling and the \pIII star IMF that
play a crucial role in the reionization process. Assuming that IMF a
power-law, the conditions that the metallicity of ionized regions must
be high enough for triggering sustained galaxy formation and \pIII
stars massive enough for seeding MBHs, greatly constrain those two
properties.

The only acceptable solutions we find are in two narrow sets, one for
small $M\III\lo$ values (moderately top-heavy \pIII star IMFs) leading
to single reionization, and another one for large $M\III\lo$ values
(top-heavier \pIII star IMFs) leading to double reionization. The
latter solutions have a first ionization phase driven by Population
III stars completed at $z\sim 10$, and after a short recombination
period a second ionization phase driven by normal galaxies completed
at $z\sim 6$. In the former solutions with single reionization, both
kinds of luminous sources contribute in parallel to the ionization of
the IGM, completed at $z\sim 6$.

The best solution with double reionization gives excellent fits to all
the observables. Instead, the best solution with single reionization
predicts a too monotonous SFH, and a slightly too steep CGH at $z\sim
4$. However, the CMB Thomson optical depth found in double
reionization ($\tau=0.102^{+0.001}_{-0.001}$), though consistent with
the observational estimates from the {\it WMAP} 9-year and {\it
  Planck} three-year data, is $3\sigma$ larger than the most recent
estimate ($\tau=0.058\pm 0.014$) from the Planck data taking into
account the large-scale polarization anisotropies. Instead, the CMB
optical depth found in single reionization
($\tau=0.072^{+0.004}_{-0.005}$) is consistent with that
estimate. There is thus some tension between the new estimate of
$\tau\es$ and the observed properties of the Universe at
high-$z$. However, the empirical values of the CMB optical depth
derived under the instantaneous reionization approximation must be
taken with caution.

The fact that with only 9 degrees of freedom it is possible to fit a
dozen of independent cosmic histories with non-trivial features
(extreme and saddle points) confers strong reliability to our
self-consistent model of the EoR.

As a byproduct, we have clarified the origin of some interesting
features shown by the data on the high-$z$ Universe: 1) the very small
sizes of spheroids at $z\sim 2-3$; 2) the sudden drop of the SFH at $z
<2$; 3) the similar behavior of the UV (and X-ray) emissivity from AGN
at the same redshift; 4) the plateau in the CGH between at $1< z < 5$;
and 5) the rough constancy of the MBH-to-spheroid mass ratio. Our
results also suggest that globular clusters are but the relics of the
lowest mass galaxies with primordial origin or, more probably, formed
during recombination in model D, which soon become frozen as satellite
galaxies.

Finally, we have provided several predictions that should allow one to
confirm the validity of the present model, and tell between single and
double reionization. The most interesting one is that, in the case of
double reionization, LAEs should be detectable near $z=10$. This
could explain the recent detection (to be confirmed) of a galaxy with
\lya\ emission at $z=8.68$ \citep{Zi15}.

\begin{acknowledgments}
Funding for this work was provided by the Spanish MINECO under
projects AYA2012-39168-C03-1, AYA2012-39168-C03-2, AYA2013-446724-P,
MDM-2014-0369 of ICCUB (Unidad de Excelencia `Mar\'ia de Maeztu'),
AYA2015-70498-C02-2-R (co-funded with FEDER funds), and the Catalan
DEC grant 2014SGR86. JMMH was supported by Spanish MINECO grant
ESP2014-59789-P.
\end{acknowledgments}

\end{document}